\documentclass[sigconf]{acmart}

\usepackage{algorithm}
\usepackage{color}
\usepackage{xcolor}
\usepackage{url}
\usepackage{subfigure}
\usepackage{xspace}
\usepackage{enumerate}
\usepackage{multirow}
\usepackage{balance}
\usepackage{epstopdf}
\usepackage{url}
\usepackage{algorithmicx}
\usepackage{algpseudocode}
\usepackage{diagbox}
\usepackage{outlines}
\usepackage{hyperref}
\usepackage[capitalize]{cleveref}

\newcommand{\squishlist}{
   \begin{list}{$\bullet$}
    {
      \setlength{\itemsep}{0pt}
      \setlength{\parsep}{3pt}
      \setlength{\topsep}{3pt}
      \setlength{\partopsep}{0pt}
      \setlength{\leftmargin}{1.5em}
      \setlength{\labelwidth}{1em}
      \setlength{\labelsep}{0.5em} } }

\newcommand{\squishend}{
    \end{list}  }

\makeatletter
\def\mathcolor#1#{\@mathcolor{#1}}
\def\@mathcolor#1#2#3{%
  \protect\leavevmode
  \begingroup
    \color#1{#2}#3%
  \endgroup
}
\makeatother

\newcommand{\eat}[1]{}

\newcommand{\stitle}[1]{\vspace{2mm} \noindent{\bf #1}}

\begin{document}

\newtheorem{remark}{Remark}

\newcommand{\op}[1]{{\sc #1}\xspace}
\newcommand{\mybullet}{\vspace{1mm}\noindent$\bullet$~~}

\newcommand{\abowd}[1]{[[\emph{\color{blue}Abowd: #1}]]}
\renewcommand{\abowd}[1]{}
\newcommand{\am}[1]{[[\emph{\color{blue}AM: #1}]]}
\newcommand{\dap}[1]{[[\emph{\color{purple}DP: #1}]]}
\newcommand{\mh}[1]{[[\emph{\color{magenta}MH: #1}]]}
\newcommand{\gm}[1]{[[\emph{\color{red}GM: #1}]]}
\newcommand{\ry}[1]{[[\emph{\color{green}RM: #1}]]}
\newcommand{\ik}[1]{[[\emph{\color{orange}IK: #1}]]}
\newcommand{\amref}[2]{{\color{cyan} #1}~[[\emph{{\color{blue} AM: #2}}]]}
\newcommand{\mhref}[2]{{\color{cyan} #1}~[[\emph{{\color{magenta} MH: #2}}]]}
\newcommand{\gmref}[2]{{\color{cyan} #1}~[[\emph{{\color{red} GM: #2}}]]}
\newcommand{\ryref}[2]{{\color{cyan} #1}~[[\emph{{\color{green} RM: #2}}]]}
\newcommand{\st}[1]{[[\emph{\color{violet}ST: #1}]]}
\newcommand{\shref}[2]{{\color{cyan} #1}~[[\emph{{\color{violet} ST: #2}}]]}

\newcommand{\planref}[1]{\textnormal{\textsc{Plan \##1}}}

\def\planidentity{{1}\xspace}
\def\planprivelet{{2}\xspace}
\def\planh{{3}\xspace}
\def\planhb{{4}\xspace}
\def\planmwem{{7}\xspace}
\def\planahp{{8}\xspace}
\def\plandawa{{9}\xspace}
\def\plandawastripe{{13}\xspace}
\def\planhbstripe{{14}\xspace}
\def\planprivbayesls{{15}\xspace}
\def\planmwemb{{16}\xspace}
\def\planmwemc{{17}\xspace}
\def\planmwemd{{18}\xspace}



\def\transform{{\sc Transform}\xspace}
\def\measure{{\sc Measure}\xspace}
\def\public{{\sc Public}\xspace}
\newcommand{\pk}{protected kernel\xspace}

\def\sys{{\normalsize$\epsilon$}{\sc ktelo}\xspace}

\newcommand{\algoname}[1]{\textnormal{\textsc{#1}}}

\newcommand{\Lone}[1]{\left\Vert #1  \right\Vert_1}
\newcommand{\norm}[1]{\left\lVert#1\right\rVert}
\newcommand{\set}[1]{\{#1\}}   

\def\C{\varmathbb{C}}
\def\api{\textrm{API}}
\def\init{\textrm{\bf Init}}
\def\Q{M}   
\def\R{R}   
\def\qsize{m}
\def\wsize{m}
\def\rsize{r}
\newcommand{\vect}[1]{\mathbf{#1}}

\def\q{\vect{q}}
\def\x{\vect{x}}

 \def\db{I}  
 \def\nbrs{nbrs}

 \def\algG{\mathcal{A}}  

\acmYear{2020}\copyrightyear{2020}
\setcopyright{acmcopyright}
\acmConference[FAT* '20]{Conference on Fairness, Accountability, and Transparency}{January 27--30, 2020}{Barcelona, Spain}
\acmBooktitle{Conference on Fairness, Accountability, and Transparency (FAT* '20), January 27--30, 2020, Barcelona, Spain}
\acmPrice{15.00}
\acmDOI{10.1145/3351095.3372872}
\acmISBN{978-1-4503-6936-7/20/01}

\author{David Pujol}
\email{david.pujol@duke.edu}
\affiliation{
	\institution{Duke University}
}

\author{Ryan McKenna}
\email{rmckenna@cs.umass.edu}
\affiliation{
	\institution{University of Massachusetts, Amherst}
}
\author{Satya Kuppam}
\email{skuppam@cs.umass.edu}
\affiliation{
	\institution{University of Massachusetts, Amherst}
}
\author{Michael Hay}
\email{mhay@colgate.edu}
\affiliation{
	\institution{Colgate University}
}
\author{Ashwin Machanavajjhala}
\email{ashwin@cs.duke.edu}
\affiliation{
	\institution{Duke University}
}
\author{Gerome Miklau}
\email{miklau@cs.umass.edu}
\affiliation{
	\institution{University of Massachusetts, Amherst}
}



\title{Fair Decision Making Using Privacy-Protected Data}

\begin{abstract}
Data collected about individuals is regularly used to make decisions that impact those same individuals.  We consider settings where sensitive personal data is used to decide who will receive resources or benefits. While it is well known that there is a tradeoff between protecting privacy and the accuracy of decisions, we initiate a first-of-its-kind study into the impact of formally private mechanisms (based on differential privacy) on fair and equitable decision-making.  We empirically investigate novel tradeoffs on two real-world decisions made using U.S. Census data (allocation of federal funds and assignment of voting rights benefits) as well as a classic apportionment problem. 

Our results show that if decisions are made using an $\epsilon$-differentially private version of the data, under strict privacy constraints (smaller $\epsilon$), the noise added to achieve privacy may disproportionately impact some groups over others. 
We propose novel measures of fairness in the context of randomized differentially private algorithms and identify a range of causes of outcome disparities. We also explore improved algorithms to remedy the unfairness observed.
\end{abstract}
\settopmatter{printfolios=true}
\maketitle
\vspace{-1em}
\section{Introduction}
Data collected about individuals is regularly used to make decisions that impact those same individuals. One of our main motivations is the practice of statistical agencies (e.g. the U.S. Census Bureau) which publicly release statistics about groups of individuals that are then used as input to a number of critical civic decision-making procedures. The resulting decisions can have significant impacts on individual welfare or political representation.  For example:

\squishlist
\item election materials must be printed in minority languages in specified electoral jurisdictions (only) if certain conditions are met, which are determined by published counts of minority language speakers and their illiteracy rates.
\item annual funds to assist disadvantaged children are allocated to school districts, determined by published counts of the number of eligible school-age children meeting financial need criteria;
\item seats in legislative bodies (national and state legislatures and municipal boards) are \textit{apportioned} to regions based on their count of residents. For example, seats in the Indian parliament are allocated to states in proportion to their population.
\squishend

In many cases, the statistics used to make these decisions are sensitive and their confidentiality is strictly regulated by law. For instance, in the U.S., census data is regulated under Title 13 \cite{censusTitle13}, which requires that no individual be identified from any data released by the Census Bureau, and data released about students is regulated under FERPA\footnote{The Family Educational Rights and Privacy Act (FERPA) 20 U.S.C. $\S$ 1232g (2012)}. In the EU, data releases are strictly regulated under GDPR\footnote{General Data Protection Regulation, Council Regulation (EU) 2016/679, art. 4, of the European Parliament}. Statistical agencies worldwide uphold privacy and confidentiality requirements by releasing statistics that have passed through a \textit{privacy mechanism}.  In the U.S., a handful of critical decisions (e.g. congressional apportionment) are made on unprotected true values, but the vast majority of decisions are made using privatized releases. Our focus is the impact of mechanisms satisfying formal privacy guarantees (based on differential privacy \cite{Dwork06Calibrating}) on resource allocation decisions.

The accuracy of the above decisions is clearly important, but it conflicts with the need to protect individuals from the potential harms of privacy breaches.  To achieve formal privacy protection, some error must be introduced into the properties of groups (i.e. states, voting districts, school districts), potentially distorting the decisions that are made.  In the examples above, the consequences of error can be serious: seats in parliament could be gained or lost, impacting the degree of representation of a state's citizens; funding may not reach eligible children; or a district deserving minority voting support may not get it, disenfranchising a group of voters.

%

The tradeoff between privacy protection and the accuracy of decision making must therefore be carefully considered.  The right balance is an important social choice \cite{Abowd18Economic}  and the model of differential privacy allows for a more precise analysis of this choice. Maximizing the accuracy achievable under differentially privacy has been a major focus of recent privacy research, resulting in many sophisticated algorithmic techniques \cite{Dwork14Algorithmic,Hay16Principled}.  Yet that effort has considered accuracy almost exclusively through aggregate measures of expected error, which can hide disparate effects on individuals or groups.



In this paper we look beyond the classic tradeoff between privacy and error to consider fair treatment in decision problems based on private data.  If we accept that privacy protection will require some degree of error in decision making, does that error impact groups or individuals equally? Or are some populations systematically disadvantaged as a result of privacy technology?  These questions are especially important now: the adoption of differential privacy is growing~\cite{Greenberg16Apples,Erlingsson14Rappor:,machanavajjhala08onthemap,haney17:census}, and, in particular, the U.S. Census Bureau is currently designing differentially private methods planned for use in protecting 2020 census data~\cite{census-url,Vilhuber17Proceedings,census2010DemonstrationDataProducts}.



The contributions of our work include the following.  We present a novel study of the impact of common privacy algorithms on the equitable treatment of individuals. In settings where the noise from the privacy algorithm is modest relative to the statistics underlying a decision, impacts may be negligible.  But when stricter privacy (i.e., small values of $\epsilon$) is adopted, or decisions involve small populations, significant inequities can arise. We demonstrate the importance of these impacts by simulating three real-world decisions made using sensitive public data: the assignment of voting rights benefits, the allocation of federal funds, and parliamentary apportionment. 

\squishlist
	\item We show that even if privacy mechanisms add equivalent noise to independent populations, significant disparities in outcomes can nevertheless result. For instance, in the federal funds allocation use case, under strict privacy settings of $\epsilon=10^{-3}$, some districts receive over 500$\times$ their proportional share of funds while others receive less than half their proportional share. Under weaker privacy settings ($\epsilon = 10$), this disparity is still observed but on a much smaller scale.

	\item For assigning voting rights benefits to minority language communities, we find that noise for privacy can lead to significant disparities in the rates of correct identification of those deserving the benefits, especially under stricter privacy settings.

	\item For the parliamentary apportionment problem, surprisingly, there are settings of $\epsilon$ where the apportionment of seats to Indian states based on the noisy data is {\em more} equitable, \emph{ex ante}, than the standard deterministic apportionment.
	\item For funds allocation and voting benefits (the allocation problems with the greatest disparities) we propose methods to remedy inequity, which can be implemented without modifying the private release mechanism. 

\squishend
Our study reveals that the use of privacy algorithms involves complex tradeoffs  which can impact social welfare.  Further, these impacts are not easy to predict or control because they may be caused by features of the privacy algorithm, the structure of the decision problem, and/or properties of the input data.  We believe these findings call for new standards in the design and evaluation of the privacy algorithms that are starting to be deployed by companies and statistical agencies.

The organization of the paper is as follows. In the next section we describe our problem setting, followed by related work in \Cref{sec:related}.  In \Cref{sec:title1,sec:vra,sec:apportionment} we investigate fairness in the example problem domains of voting rights, funds allocation, and apportionment, respectively.  We conclude with open challenges in \Cref{sec:conclusion}.  The appendix includes algorithm details to aid reproducibility, and proofs, but is not essential to the claims of the paper.

\stitle{Remark:}
{\em This work uses only public data, released by the U.S. Census Bureau and other institutions.
Our empirical results do not measure the actual impacts of any agency practice currently in use.  Instead, we {\em simulate} the use of state-of-the-art privacy algorithms on real use-cases in order to understand and quantify {\em potential} unfair impacts, should these privacy algorithms be adopted.}

\section{Problem Setting} \label{sec:background}

Below we provide a general definition of the assignment problems we consider, define differential privacy and assignment based on private inputs, as well as our methodology for assessing fairness.
\subsection{Assignment Problems}
 We assume a universe of individuals each described by a record in a table $I$.  Individuals are divided into disjoint {\em assignee populations}, each population denoted by a label $ a \in A$. In our example problems, assignee populations are characterized by, and labeled with, the geographic region in which they reside (e.g. state, county, school district).  For example, we may have $a=\mbox{Wyoming}$ and use $I_a$ to denote the set of records for all Wyoming residents.

An assignment method $\mathcal{M}:A \rightarrow O$ associates a resource or benefit with each assignee population, formalized as an {\em outcome} from an outcome set $O$.  We are primarily concerned with equitable treatment of assignee populations in terms of the outcomes they receive from an assignment.

The assignment methods we consider are deterministic (in the absence of privacy protection) and depend on properties of the assignee populations, which are described by statistics. These are formalized by one or more statistical queries $Q$, evaluated on the records corresponding to the assignee population. For example, we may write $Q=\{ tot \}$ where $tot(I_a)$ is the query that computes the total population of an assignee $a$.
These statistics are stored in a matrix $\mathbf{X} \in \mathbb{R}^{A \times Q}$, indexed by elements $a \in A$ and $q \in Q$ such that $ \mathbf{X}_a^q = q(I_a)$.
An assignment method $\mathcal{M}$ will typically be defined with respect to this matrix of statistics $\mathbf{X}$, and we use the notation $\mathcal{M}(a; \mathbf{X}) $ to make this dependence clear.  The vector of outcomes $ \mathbf{o} \in O^A $, formed from elements $ \mathbf{o}_a = \mathcal{M}(a; \mathbf{X}) $, is the {\em ground truth assignment} because it is computed on the true, unmodified statistics about the assignee populations.

\Cref{tbl:titleI,tbl:vra,tbl:apportionment} (in the later sections) contain the formal descriptions of the assignment methods for our three example problems, including a specification of the assignee population, the outcome space, the query set, and the rule underlying the assignment.
\subsection{Differential Privacy}

Differential privacy \cite{Dwork06Calibrating,Dwork14Algorithmic} is a formal model of privacy that offers each individual a persuasive guarantee: any released data computed from the sensitive input would have been almost as likely had the individual opted-out.  More formally, differential privacy is a property of a randomized algorithm that bounds the ratio of output probabilities induced by changes to an individual's data. Let $\nbrs(\db)$ be the set of databases differing from $I$ in at most one record.

\begin{definition}[Differential Privacy~\cite{Dwork06Calibrating}] \label{def:diffp}
A rand\-om\-ized algorithm $\algG$ is $\epsilon$-differentially private if for any instance $\db$, any $\db' \in \nbrs(\db)$, and any outputs $O \subseteq Range(\algG)$:
\vspace{-1em}
$$
Pr[ \algG(\db) \in O] \leq \exp(\epsilon) \times Pr[ \algG(\db') \in O]
$$
\end{definition}
%

Differentially private algorithms protect individuals and all of their associated properties, and in addition, every individual enjoys the same bound on privacy loss, which is quantified by (a function of) the privacy parameter $\epsilon$.  Smaller $\epsilon$ implies greater privacy but greater noise, and the parameter $\epsilon$ is sometimes referred to as the privacy loss ``budget''.  A useful property of statistics computed in a differentially private manner is that any subsequent computations that use those statistics are also differentially private for the same $\epsilon$ (assuming they do not also use the sensitive data).

We use two privacy mechanisms in this paper. The first is the standard Laplace mechanism~\cite{Dwork06Calibrating}.  While the Laplace Mechanism is a fundamental building block of many differentially private algorithms, it can offer sub-optimal error if applied directly to some tasks.
Therefore, we also consider the {Data- and Workload-Aware} (DAWA) algorithm \cite{Li14Data-}.  It is one of a number of recently-proposed algorithms~(cf. \cite{Hay16Principled}) which introduce complex noise that is adapted to the input data.
These techniques can offer substantially reduced error rates in some settings, but may introduce statistical bias in the estimates produced.  This is in contrast to the Laplace mechanism, which produces unbiased estimates, and with error that is independent of the input.
We chose DAWA because it was reported to perform well in benchmarks \cite{Hay16Principled}.  In each of the sections that follow we describe how these algorithms are adapted to the allocation problems studied. We provide further background in the appendix. 
\vspace{-1em}
\subsection{Assignment Using Private Inputs}
Given an assignment problem, we protect the privacy of the members of each assignee population by answering the queries in $Q$ using differentially private mechanism $\algG_Q$. The resulting noisy query answers satisfy differential privacy for a given privacy parameter $\epsilon$: $\algG_Q(I,\epsilon)=\mathbf{\tilde X}$.   We then assume the private assignments are computed with $\mathcal{M}$, using $\mathbf{\tilde X}$ in place of $\mathbf{X}$: $ \mathbf{\tilde o}_a = \mathcal{M}(a; \mathbf{\tilde X})$.  As noted above, $\mathbf{\tilde o}$ inherits the privacy guarantee of $\mathbf{\tilde X}$.

While $\mathcal{M}$ is deterministic, when $\mathcal{M}$ is composed with the randomized private computation of statistics, the result is a randomized assignment algorithm, inducing a probability distribution over outcome vectors.  Assessments of fairness must therefore be probabilistic in nature.  The expected error in the statistics, introduced by the privacy mechanism is: $\mathbb{E}[|| \mathbf{\tilde X} - \mathbf{X} ||]$ (for a suitable metric $|| \cdot ||$) which we distinguish from error in the outcome space: $\mathbb{E}[||\mathbf{\tilde o} - \mathbf{o}||] $

Note that we assume the private computation of the supporting statistics $Q$ is followed by an assignment method $\mathcal{M}$. In this initial work, we restrict our attention to this {\em data publishing} model because it follows the practice of many statistical agencies: they release fixed sets of statistics (after invoking disclosure limitation methods) which are used for a variety of purposes.  In experiments we therefore begin by measuring the effects of current practice: applying the standard assignment method to the privatized data.  But we also consider remedy approaches that alter the assignment method to account for the presence of noise introduced by the privacy mechanism. Other alternatives (namely altering the privacy mechanism itself) are noted as future work in \Cref{sec:conclusion}.

\subsection{Methodology}
The example problems we consider in \Cref{sec:title1,sec:vra,sec:apportionment} assign resources or benefits to populations according to properties of those populations that define their {\em entitlement}.  For example, for Title 1 funding (\Cref{sec:title1}), a school district's entitlement is proportional to the number of students who meet a specific qualification condition.  Our goal is not to question the fairness of the declared entitlement or the resulting ground truth assignment, as these are typically mandated by law.  Instead, we consider the change in outcomes due to the introduction of privacy protection.

Since different populations have different entitlements, we do not seek to treat each population equally, but instead to treat equals equally.  However, with a randomized assignment method, even identical populations will receive different outcomes over runs of the algorithm, so we must evaluate equal treatment in expectation or with high probability.  We provide problem-specific fairness measures in the following sections.


\section{Related Work} \label{sec:related}

While fairness and privacy are topics that have been considered by philosophers and theologians for thousands of years, it is only recently that these values have begun to be engineered into algorithms.  Differential privacy \cite{Dwork14Algorithmic,Dwork06Calibrating} provides a formal model for reasoning about and controlling a quantitative measure of privacy loss.  Fairness has been formalized in economics, and, more recently, in definitions emerging from machine learning \cite{Dwork12Fairness,Verma18Fairness,Kleinberg18Inherent,Chouldechova17Fair,Romei13multidisciplinary}.

Yet relatively little work has considered the direct interaction of privacy and fairness.   Dwork and Mulligan \cite{Dwork13Privacy} warn against the expectation that privacy controls and transparency alone can offer resistance to discrimination in the context of large-scale data collection and automated classification.  And Dwork et al. \cite{Dwork12Fairness} propose a framework for fair classification which they show can be viewed as a generalization of differential privacy.  Both of these focus on settings distinct from ours.  Conceptually closest to our work is a recent position paper in which Ekstrand et al. \cite{Ekstrand18Privacy} raise a number of questions about the equitable provision of privacy protections (privacy-fairness) and equitable impacts of privacy-protection mechanisms (accuracy-fairness).  In addition, very recently, Bagdasaryan and Shmatikov \cite{bagdasaryan_shmatikov_2019} have shown the disparate impact of differential privacy on learned models. 



\stitle{Economics and Social Choice Theory:}
Abowd and Schmutte~\cite{Abowd18Economic} characterize accuracy and privacy protection as competing social goods and invoke an economic framework in which the demand for accuracy is balanced with the demand for privacy.  They use the model of differential privacy to quantify privacy loss and study Title I funds allocation in detail.

They measure inaccuracy using total squared error, a standard metric in the privacy community, and explain that this corresponds to utilitarian social welfare.  This work inspired ours, motivating us to ask whether there are other social welfare functions to consider in the design of privacy algorithms.

In the literature on social choice, fair allocation methods have been widely studied.  Two of the example problems we consider are instances of fair division problems.  Funds allocation is a fair division problem for a divisible and homogeneous resource (since money can be divided and only the amount matters) where agents (in our example, school districts) value the resource equally but have differing rights to the resource (e.g. based on eligible population).  This is a trivial fair division problem whose solution is a proportional division.  In our setting, the division deviates from proportional because the information about agents' rights to the resource is noisy.  We are not aware of fairness definitions which consider this variant directly, although Xue \cite{Xue18Fair} considers a related scenario where agents rights are uncertain and proposes a division that discounts an agent's allocation accordingly.

Apportionment is a fair division problem for an indivisible and homogeneous good (since seats cannot be divided and only the number of seats matters) where agents (in our example, states) value the resource equally but have differing rights to the resource (determined by population).  Again, in our setting we must consider the impact of noisy information about agents' rights.  While the study of apportionment methods and their properties has a long history, we are aware of no existing approaches to cope with noisy inputs.  The closest related work may be that of Grimmett \cite{Grimmett04Stochastic} which proposes a randomized apportionment method along with a fairness criterion we consider in \Cref{sec:apportionment}.

\stitle{Fairness in Machine Learning:}
A number of fairness definitions have been proposed recently for assessing the impacts of predictive algorithms \cite{Dwork12Fairness,Verma18Fairness,Kleinberg18Inherent,Chouldechova17Fair,Romei13multidisciplinary}, primarily focused on algorithms that assign scores or classifications to individuals.  Fairness criteria measure the degree of disparate treatment for groups of individuals who should be treated equally (e.g. males and females in the context of hiring).  Our example problem concerning minority language benefits is related since the goal is to classify jurisdictions.  However, rather than studying the impact of a classifier that may display biased performance on unseen examples, we have a fixed decision rule (mandated by law) but error is introduced into outcomes because of noise in the input statistics.  Although we could certainly compare impacts across groups (e.g. whether Hispanic and Chinese minority language speakers are treated equally) we are also concerned with equitable treatment of arbitrary pairs of jurisdictions.  The metric we use for this problem is related to error rate balance \cite{Chouldechova17Fair} but other metrics could also be considered.

\vspace{-0.1em}
\stitle{Statistical Agency Practices:}
Statistical agencies like the U.S. Census Bureau have considered the potential impacts of inaccuracy and bias in their data products for decades.  Broadly, errors may arise from sampling, data cleaning, or privacy protection.  Census data products derived from surveys (rather than censuses) include margins-of-error representing estimates of uncertainty due to sampling.  Margins-of-error are intended to quantify sampling error but have {\em not} historically considered the distortion introduced by the data transformations applied for privacy protection.

In most cases, released data are treated as true by end users: assignment and allocation methods are applied directly to released summary statistics without any modification to take into account potential inaccuracies.  We are not aware of systematic studies of potential bias in the statistics currently released by statistical agencies, however Spielman observed that margins-of-error can be correlated with income levels in some Census products, leading to greater inaccuracies for low-income persons~\cite{Spielman15Reducing,Spielman14Patterns}.

The Census Bureau will be adopting differential privacy for parts of the 2020 Census of Population and Housing \cite{census-url}. This motivates a careful consideration of the implications of differentially private mechanisms on both accuracy and fairness. A preliminary version of the planned algorithm was released by the Census Bureau subsequent to this work~\cite{census2010DemonstrationDataProducts}.

While the U.S. Census Bureau is required by law to protect individuals' privacy, it  is also obligated to support accurate decision making. It therefore makes strategic choices about the accuracy of its released products. For some critical assignment problems (e.g. apportionment and redistricting), the Census forgoes privacy protection in order to favor accurate allocation and this choice is supported by law.  In other cases, such as minority language benefits, special variance reduction methods have been adopted to boost accuracy \cite{vra-stat-methodology}.  Ultimately, for legacy privacy methods employed by the Census, it is not possible for users to evaluate potential biases.

\section{Problem 1: Minority Language Voting Rights} \label{sec:vra}

\newcommand{\ident}{D-Laplace\xspace}

\begin{figure*}[t]
	\centering
	\subfigure[The \ident algorithm \eat{, $\epsilon=\{.01, .1, 1.0, 10.0\}$}]{
	\includegraphics[width=0.24\textwidth]{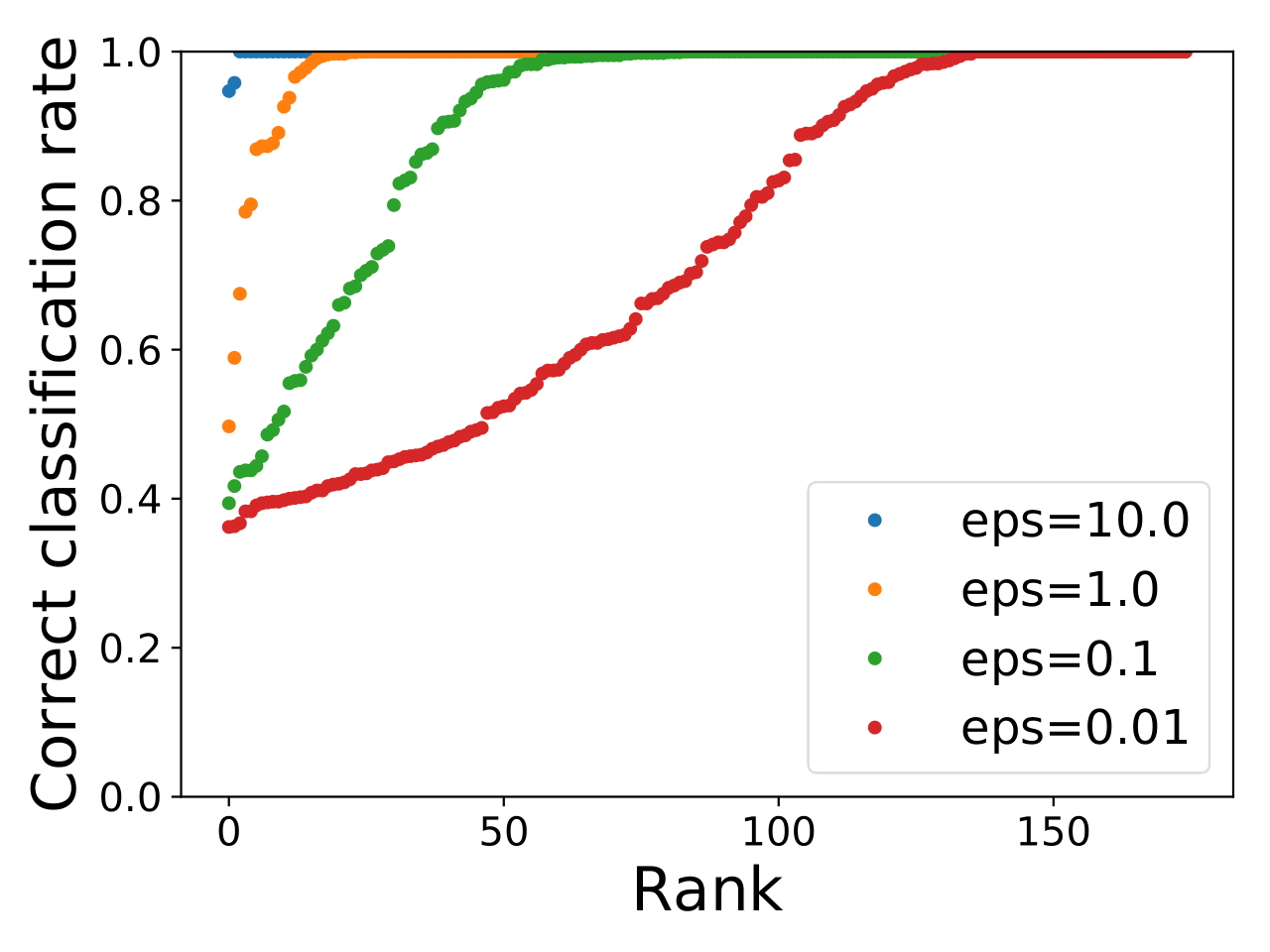}
	\label{fig:vra-ident}}
	\subfigure[The DAWA algorithm \eat{,$\epsilon=\{.01, .1, 1.0, 10.0\}$}]{
	\includegraphics[width=0.24\textwidth]{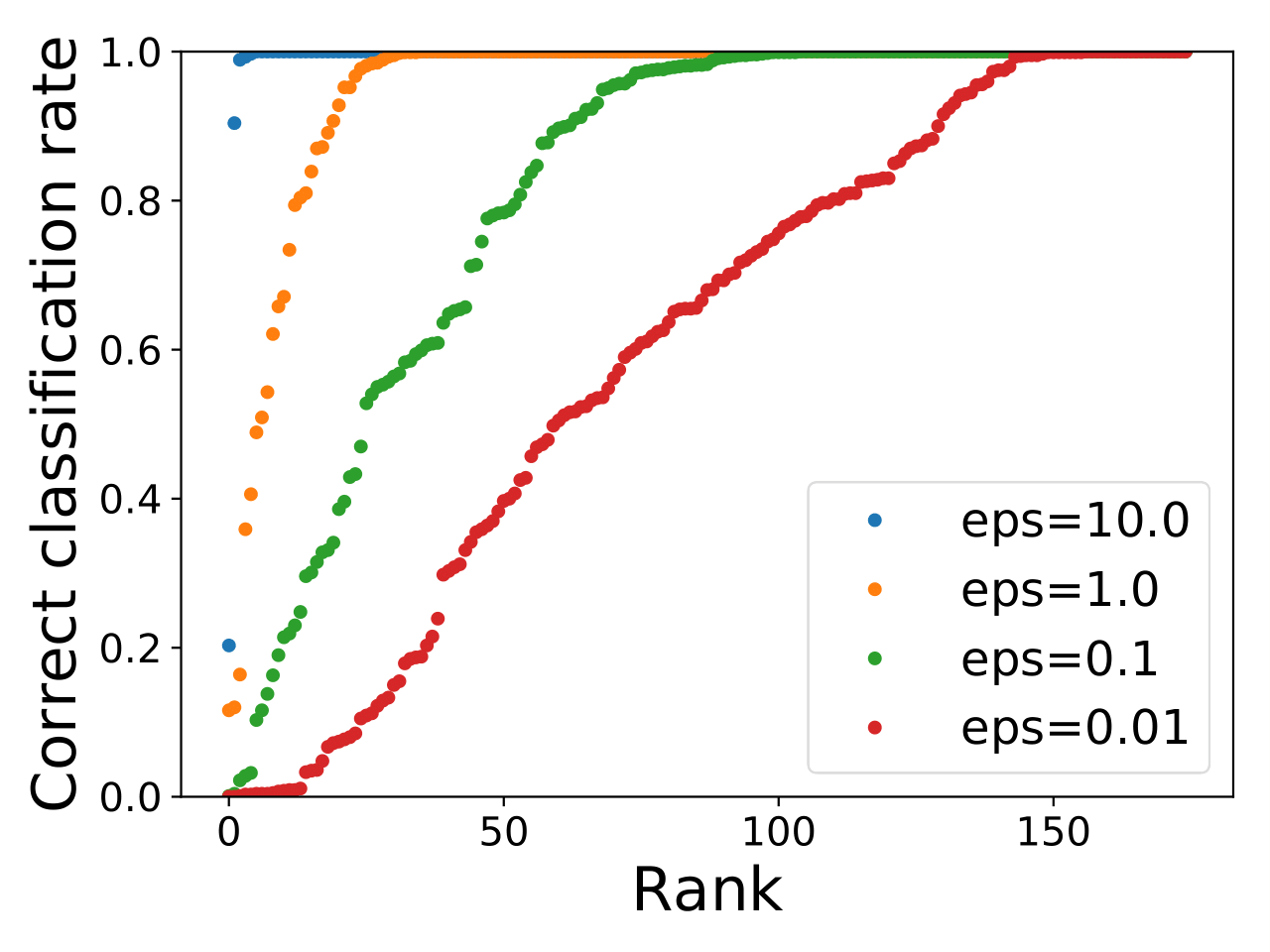}\label{fig:vra-dawa}}
	\subfigure[distance to threshold, $\epsilon=.1$]{
	\includegraphics[width=0.24\textwidth]{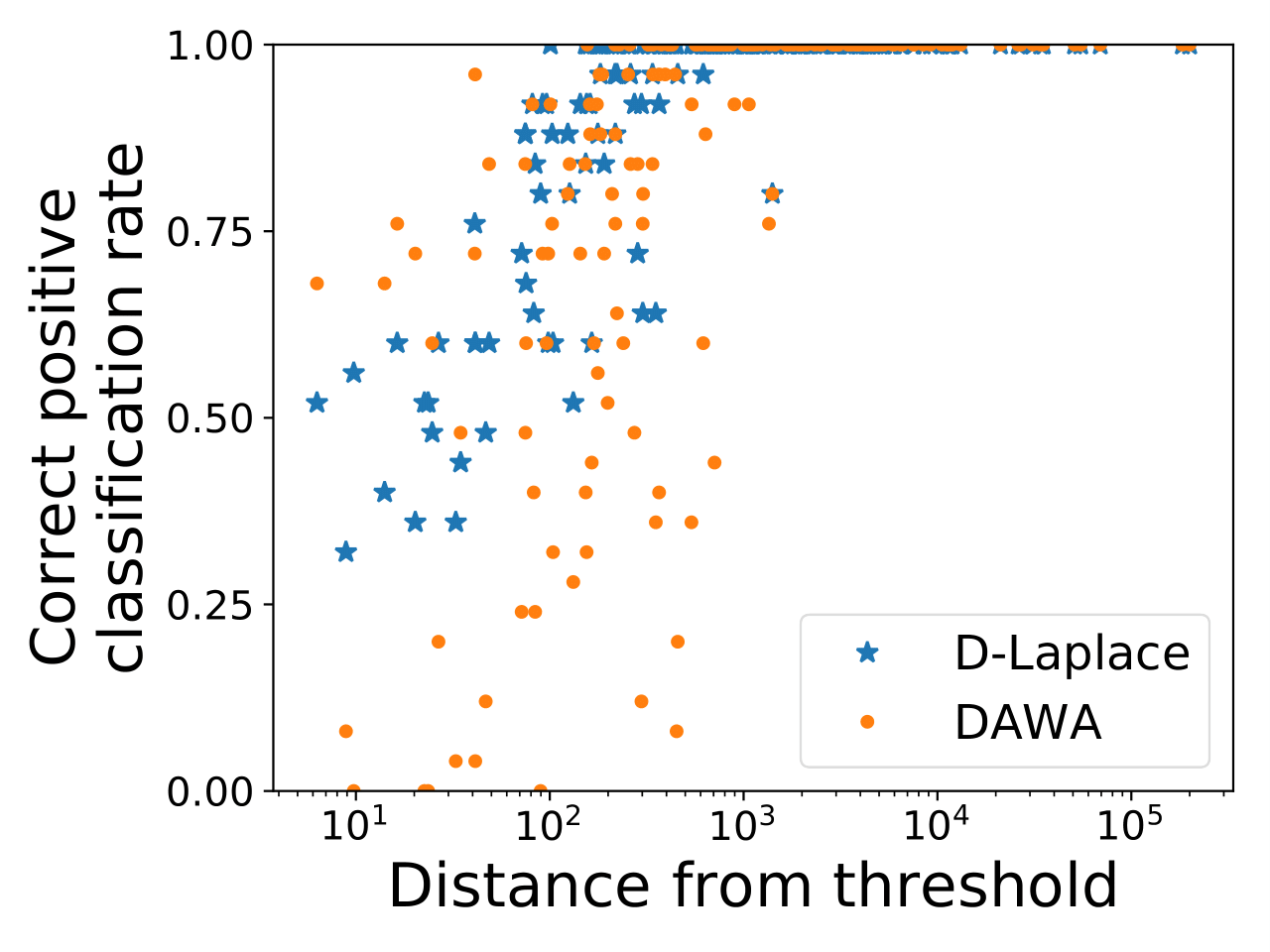}\label{fig:vra-dist}}
	\subfigure[Repair mechanism, for varying $p$ and $\epsilon$.]{
	\includegraphics[width=0.24\textwidth]{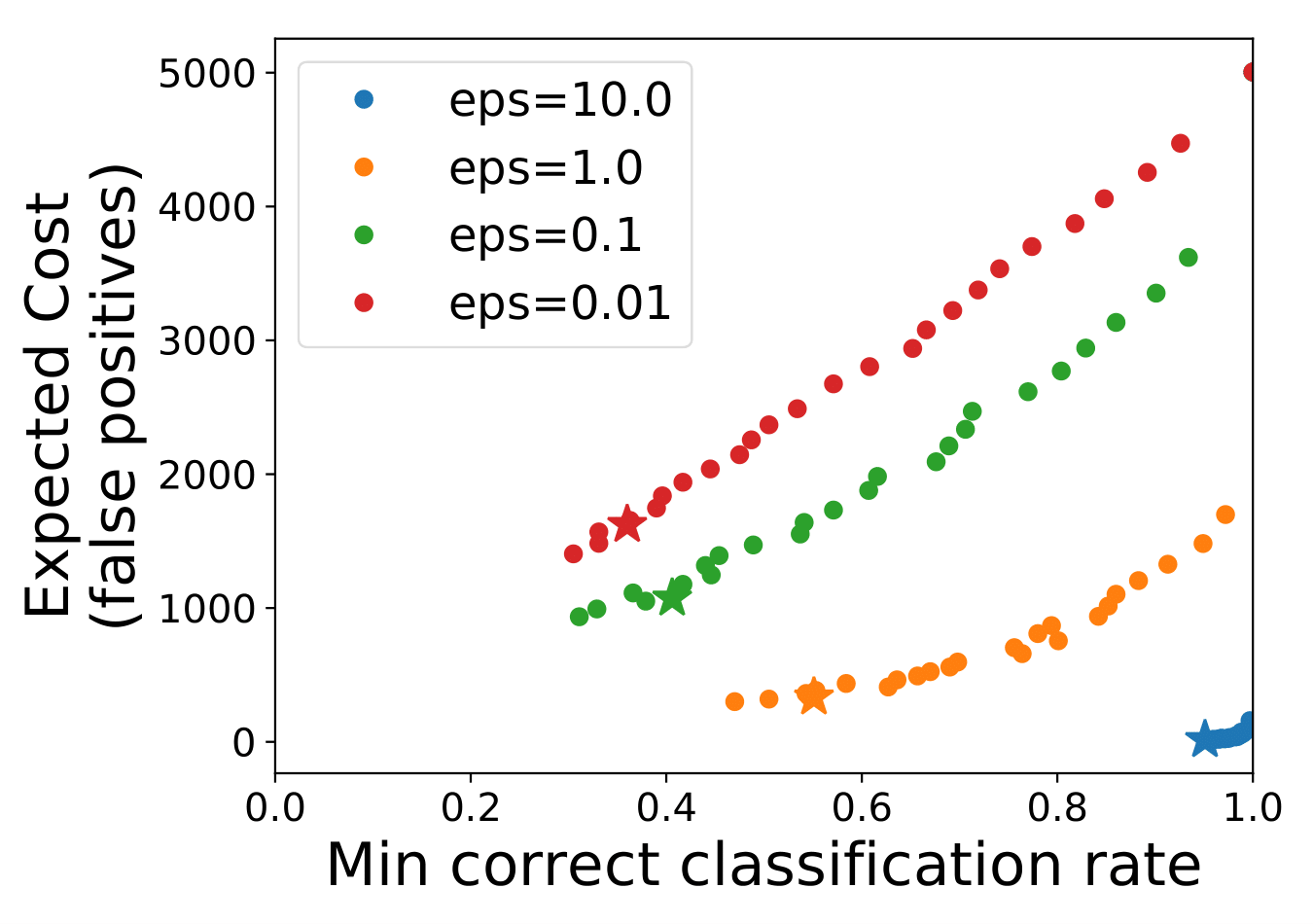}\label{fig:cost}}
\vspace{-0.5em}
\caption{\label{fig:vra} Minority Language Determinations using \ident and DAWA. }
\vspace{-0.5em}
\end{figure*}

The Voting Rights Act is federal legislation, passed in 1965, which provides a range of protections for racial and language minorities.  Among its many provisions is Section 203, describing conditions under which local jurisdictions must provide language assistance during elections.  Each jurisdiction (e.g. a county) is evaluated for each of 68 identified minority languages. If they meet the conditions, they are found to be ``covered'' by the provision, and must provide all election information (including voter registration, ballots, and instructions) in the minority language.

The coverage determination is made by the Census Bureau every five years, using published population statistics.  Most recently, in 2016, 263 jurisdictions (out of a total of approximately 8000) were found to be covered under Section 203, across all language minority groups \cite{vra-determination-file}.  While a small fraction of all jurisdictions are covered, an estimated 21.7 million voting-age citizens lived in these jurisdictions and were potentially impacted by this benefit.
\subsection{Problem Definition}

Informally, a jurisdiction is covered for a language if it (i) has a large enough population of voting age citizens who speak the language and have limited proficiency in English, and (ii) if the illiteracy rate of those speaking the language is higher than the national average. Condition (i) can be satisfied in either of two ways: in percentage terms ($>5\%$) or absolute terms ($>10,000$).
\Cref{tbl:vra} formalizes these criteria, defining a binary outcome (``covered'' or ``not-covered'') for each jurisdiction and for each minority language category.
\abowd{I hope you asked Robert Ashmead to share the final version of the technical paper with you. He wrote most of it. I gave instructions to have it posted on the language determinations page, but it's not there. Attached to this reply.}

\begin{table}[h]
\caption{\label{tbl:vra} Voting Rights, Minority Language Determinations}
\vspace{-2ex}
\fbox{\begin{minipage}{\columnwidth} {\small
Assignees are all combinations of U.S. voting jurisdictions with each of 68 minority language categories.
\begin{outline} \itemsep .5ex
\1 	Assignees: $a = (j,l) \in \mbox{Jurisdictions} \times \mbox{Languages}$
\1  Outcomes: $\{$Covered, Not-covered$\}$
\1 $Q = \{ vac, lep, lit \}$ where
	\2[] $vac(I_a)$: voting age citizens in $j$ speaking language $l$.
	\2[] $lep(I_a)$: voting age citizens in $j$ speaking language $l$, and limited-English proficient.
	\2[] $lit(I_a)$: voting age citizens in $j$ speaking language $l$, limited-English proficient, and less than 5th grade education.
\1 $\mathcal{M}(a; \mathbf{X}) = \Big(\frac{\mathbf{X}_a^{lep}}{\mathbf{X}_{a}^{vac}} > 0.05 \vee \mathbf{X}_a^{lep} > 10000 \Big) \wedge \frac{\mathbf{X}_a^{lit}}{\mathbf{X}_a^{lep}} > 0.0131$
\end{outline}}
\end{minipage} }
\vspace{-1em}
\end{table}


\stitle{Assessing Fairness}
To evaluate fairness we measure, for each jurisdiction, the rate of correct classification. For a covered jurisdiction $j$, i.e. $\mathcal{M}((j,l); \mathbf{X})=\mbox{`Covered'}$ where $l=\mbox{`Hispanic'}$, we measure $Pr[ \mathcal{M}(a; \mathbf{\tilde X})=Covered]$ where the probability is over randomness in the privacy algorithm.  Similarly, for a not-covered jurisdiction $j$ we measure $Pr[ \mathcal{M}(a; \mathbf{\tilde X})=\mbox{`Not-covered'}]$. We evaluate the rates of correct classification across the set of covered and not-covered jurisdictions, measuring the disparity in classification accuracy.
\subsection{Empirical Findings}
\stitle{Experimental Setup}
We use the 2016 public-use data accompanying the Census voting rights determinations, treating it as ground truth.  We focus on the ``Hispanic'' minority language group and jurisdictions that are counties or minor civil divisions.  	This data provided the values for the variables described in \cref{tbl:vra}, namely ${lep}, {vac}, {lit}$ for 5180 jurisdictions, of which 175 were Covered.


We consider two algorithms for computing the noisy statistics $\mathbf{\tilde X}$.  The first, which we call \ident, is an adaptation of the Laplace mechanism in which we decompose the original required queries $Q = \{ vac, lep, lit \}$, which together have sensitivity 3, into $Q'=\{lit, lep-lit, vac-lep\}$, which compose in parallel and have sensitivity 1.  We use the Laplace mechanism to estimate answers to $Q'$ and then derive estimates to $Q$ from them.  In our experiments this performed consistently better than a standard application of the Laplace mechanism.  The second algorithm is DAWA, as described in \cref{sec:background}, and with additional background provided in the appendix.  We run 1000 trials of each algorithm for each $\epsilon$ value.

\stitle{Finding M1:} {\em There are significant disparities in the rate of correct classification across jurisdictions.}  Because the failure to correctly classify a true positive is a more costly mistake (potentially disenfranchising a group of citizens) our results focus on the classification rate for the truly covered jurisdictions. For the 175 jurisdictions positively classified for the ``Hispanic'' language, \Cref{fig:vra-ident} shows the correct classification rate for each jurisdiction under the \ident algorithm, for four settings of the privacy parameter $\epsilon$. Jurisdictions are ranked from lowest classification rate to highest.  For $\epsilon=10.0$, all of the jurisdictions have a correct classification rate greater than 95\%. For $\epsilon=1.0$, 92\% of jurisdictions have a correct classification rate greater than 95\%, while 74\% do for $\epsilon=.1$ and 33\% do for $\epsilon=.01$. However, the plot shows that the lowest correct classification rate is about 37\% for $\epsilon=.01$ and $\epsilon=.1$ and 55\% for $\epsilon=1.0$. 



The conditions of Section 203 impose thresholds on language minority populations (as shown in \Cref{tbl:vra}).  A given covered jurisdiction may be closer to the thresholds, making it more likely that perturbation from the privacy mechanism will cause a failure to classify accurately.  As a particular example, consider Maricopa county (Arizona) and Knox county (Texas), which are both covered jurisdictions.  Maricopa county is correctly classified 100\% of the time by \ident at $\epsilon=.1$, while Knox county is correctly classified only 63\% of the time.  Because the \ident algorithm produces unbiased noise of equivalent magnitude to each jurisdiction, this difference is fully explained by the distance to the classification threshold: Maricopa county is further from the threshold than Knox county so it is more robust to the addition of noise.  Additionally, the distance to the classification threshold is strongly correlated with the population size, which is over 4,000,000 for Maricopa county, but less than 4,000 for Knox county.

Thus, in this case, the significant differences in the rate of successful classification across jurisdictions is a consequence of the decision rule and its interaction with the noise added for privacy.

Although not shown in \Cref{fig:vra}, there are also significant disparities in classification rates for the negative class (uncovered jurisdictions).
For example, the correct negative classification rate for \ident at $\epsilon = .1$ ranges from 54\% to 100\%.  Mistakes on the negative class mean, in practice, that minority language materials would be required of a jurisdiction which does not truly qualify, resulting in an unnecessary administrative and financial burden.

\stitle{Finding M2:} {\em While the DAWA algorithm offers equal or lower error on the underlying statistics for small $\epsilon$, it exacerbates disparities in classification rates.}
\Cref{fig:vra-dawa} shows a similar plot but for the DAWA algorithm, however in this case the disparities are even greater.  The lowest classification rates are zero, for both $\epsilon=.01$ and $\epsilon=.1$, implying that a few covered jurisdictions will definitely be not-covered for every run of the algorithm.  Even with higher $\epsilon$ values of $1.0$ and $10.0$, the lowest classification rates are below 25\%.  At the high end, for $\epsilon=10.0$, 99\% of the jurisdictions have a correct classification rate greater than 95\%, while 87\% do for $\epsilon=1.0$, 61\% do for $\epsilon=0.1$ and 22\% do for $\epsilon=.01$.

It is important to note that the DAWA algorithm offers approximately equivalent error on the statistics $\mathbf{X}$ compared to \ident (at $\epsilon=.1$) and in fact offers $30\%$ {\em lower} error at $\epsilon=.01$.  This is a critical finding for designers of privacy algorithms: optimizing for aggregate error on published statistics does not reliably lead to more accurate or fair outcomes for a downstream decision problem.

\stitle{Finding M3:} {\em A jurisdiction's distance from the nearest threshold explains classification rates for \ident but not DAWA.} We plot in \Cref{fig:vra-dist} a jurisdiction's euclidean distance from the nearest classification threshold against the rate of correct classification
(for $\epsilon = 0.1$).
We see that the results for \ident are well-explained: correct classification rate increases with distance from the threshold and occurs in a fairly tight band for any given distance measure.

For the DAWA algorithm, however, we observe a different result. Jurisdictions very far from the threshold have high classification rates, as expected, presumably because there is simply not enough noise to cause a failure for these cases.  But for jurisdictions a smaller distance from the threshold, there is a wide spread of classification rate and some jurisdictions reasonably far from the threshold have very low classification rates.  This shows the impact of the bias introduced in by DAWA: it sometimes groups together qualified jurisdictions with unqualified ones, causing them to be mis-classified.
\subsection{Mitigating unfairness}
We now consider the problem of modifying the allocation mechanism to alleviate some of the fairness concerns identified above.  To achieve this goal, we focus on the Laplace mechanism as the underlying mechanism, leaving its privatized counts unchanged. Rather than apply the standard allocation rule to the noisy counts, we propose an altered allocation method which can account for the noise, prioritizing correct positive classification.  In this context, this mechanism allows minimizing disenfranchised voters at the cost of unnecessarily providing voting benefits to some jurisdictions.

Given noisy counts $\mathbf{\tilde x_a}$ for jurisdiction $a$, the approach investigated above simply applies the assignment rule, returning $\mathcal{M}(a; \mathbf{\tilde x_a})$.  Instead, the principle behind our repair algorithm is to estimate the posterior probability that the jurisdiction is Covered given the observed noisy counts, i.e. we would like to estimate $Pr[\mathcal{M}(a; \mathbf{x_a})=Covered \;|\; \mathbf{\tilde x_a}]$.  We will then consider the jurisdiction covered if the estimated probability is higher than a supplied parameter $p$, which allows a tradeoff between false negatives and false positives.  With low values of $p$, most of the jurisdictions that should be covered will be, but a larger number of jurisdictions that do not deserve coverage will also be Covered.
In our implementation, we place a uniform prior distribution over the unknown quantities, which in this case are the true population counts of a jurisdiction, and estimate the probability using Monte Carlo simulation (we draw 100 samples in experiments).



In \cref{fig:cost}, we show the results of running the repair algorithm on the 2016 public-use data.  For four settings of $\epsilon$, the x-axis shows the minimum correct classification rate resulting from a range of settings of $p$.  This is a measure of disparity, since the maximum correct classification rate is always 1.0 in the cases considered.  We are thus able to reduce disparity, but must bear the cost of misclassifying jurisdictions, which is shown on the y-axis.

For modest $\epsilon=1$ this tradeoff seems appealing: while the standard algorithm (shown in the plot as a star) has a minimum correct classification rate of 0.54\% with an expected 334 false positives, we can raise the classification rate to .80 if we are willing to tolerate 870 false positives. For smaller $\epsilon$ values, the cost is greater, but the repair algorithm can allow for raising the extremely low minimum classification rates borne by some jurisdictions.  Thus, the algorithm could allow policy makers to weigh the risk of disenfranchised voters against the cost of over-supply of voting materials.

Note that cost here is expressed in terms of the expected number of jurisdictions misclassified.  Presumably the financial cost results from creating and distributing minority language voting materials in jurisdictions for which it is not legally required.  A more nuanced evaluation of cost could measure the number of individuals in those jurisdictions, since the true cost is likely to have a term that is proportional to the voting age population of the jurisdiction.

Our approach to repair has the advantage that its mitigating effects are achieved without requiring the data publisher to change their method for producing private counts.  However it does rely on the fact that the underlying noise added to counts has a known distribution.  This holds for the Laplace mechanism and some other privacy mechanisms, but does not hold for the DAWA algorithm.  Post-processing noisy outputs is a common technique used to improve the utility of privacy mechanisms. Estimating posterior distributions from differentially-private statistics has been studied previously \cite{williams2010probabilistic,bernstein2018differentially,bernstein2019differentially}.
\vspace{-0.5em}
\section{Problem 2: Title I Funds Allocation} \label{sec:title1}

\begin{figure*}[t]
	\centering
	\subfigure[Multiplicative Allocation Error]{
	\includegraphics[width=0.3\textwidth]{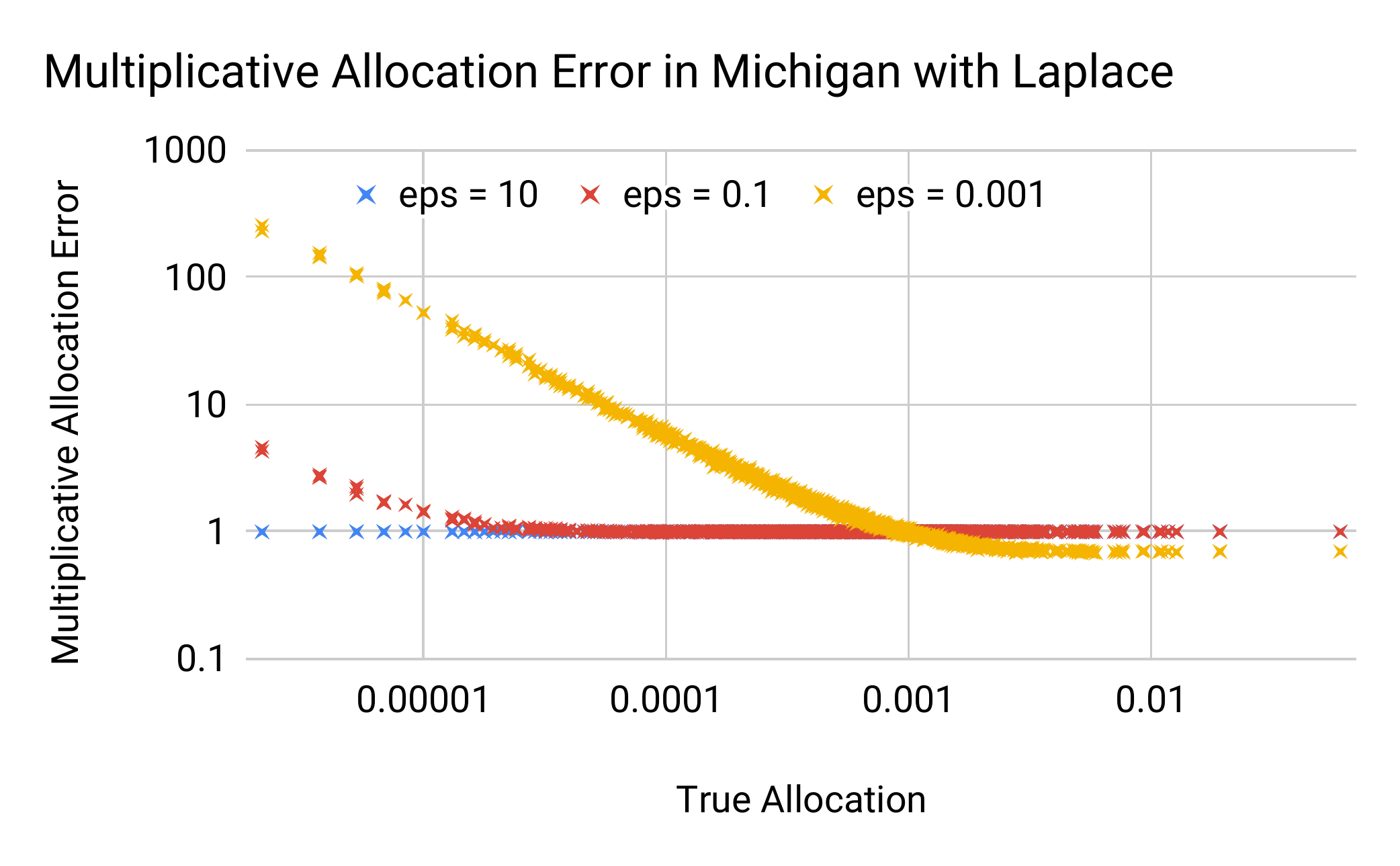}
	\label{fig:michigan-mult-lap}}
	\subfigure[Misallocation]{
	\includegraphics[width=0.3\textwidth]{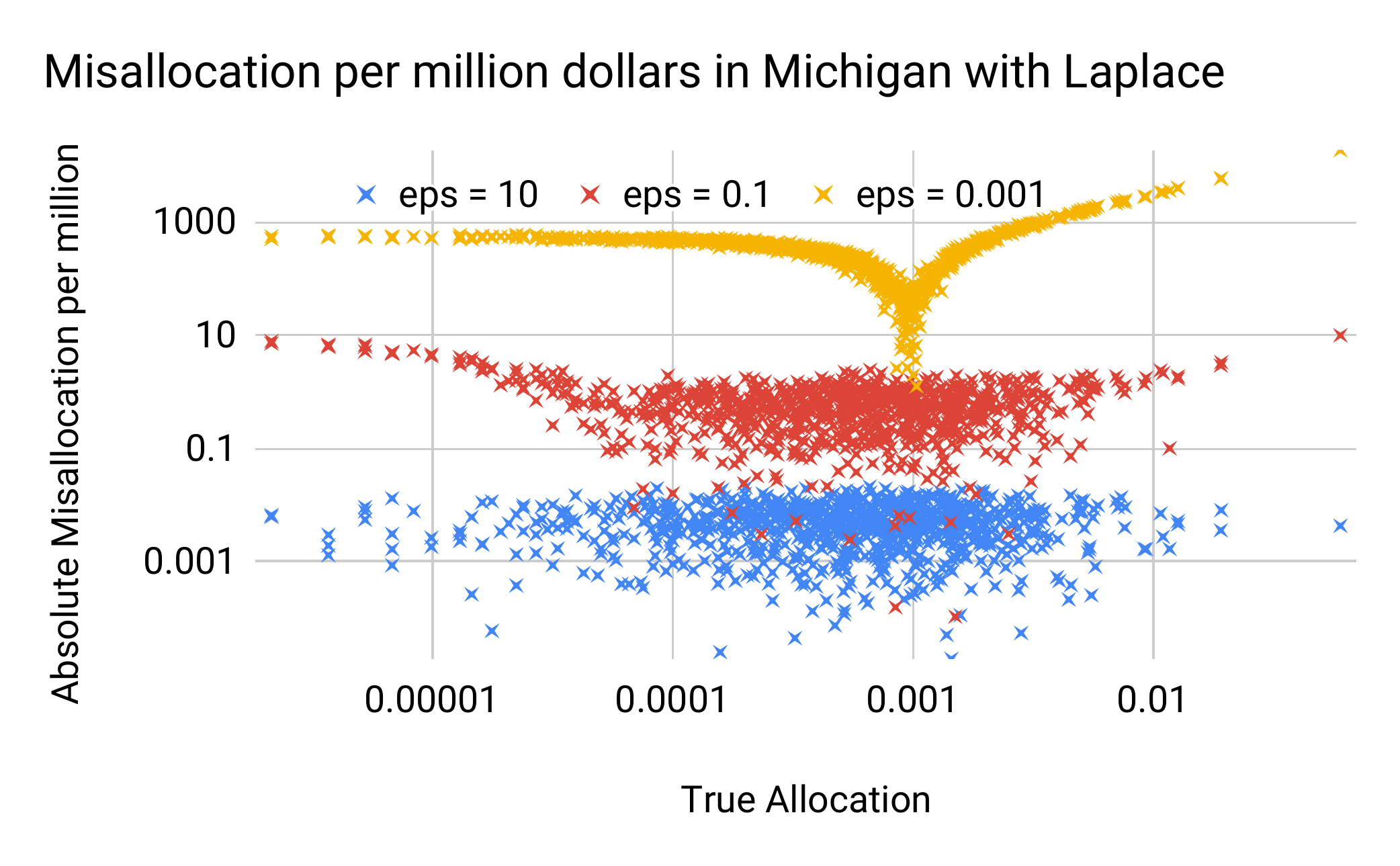}\label{fig:michigan-misallocation-lap}}
	\subfigure[Allocations]{
	\includegraphics[width=0.3\textwidth]{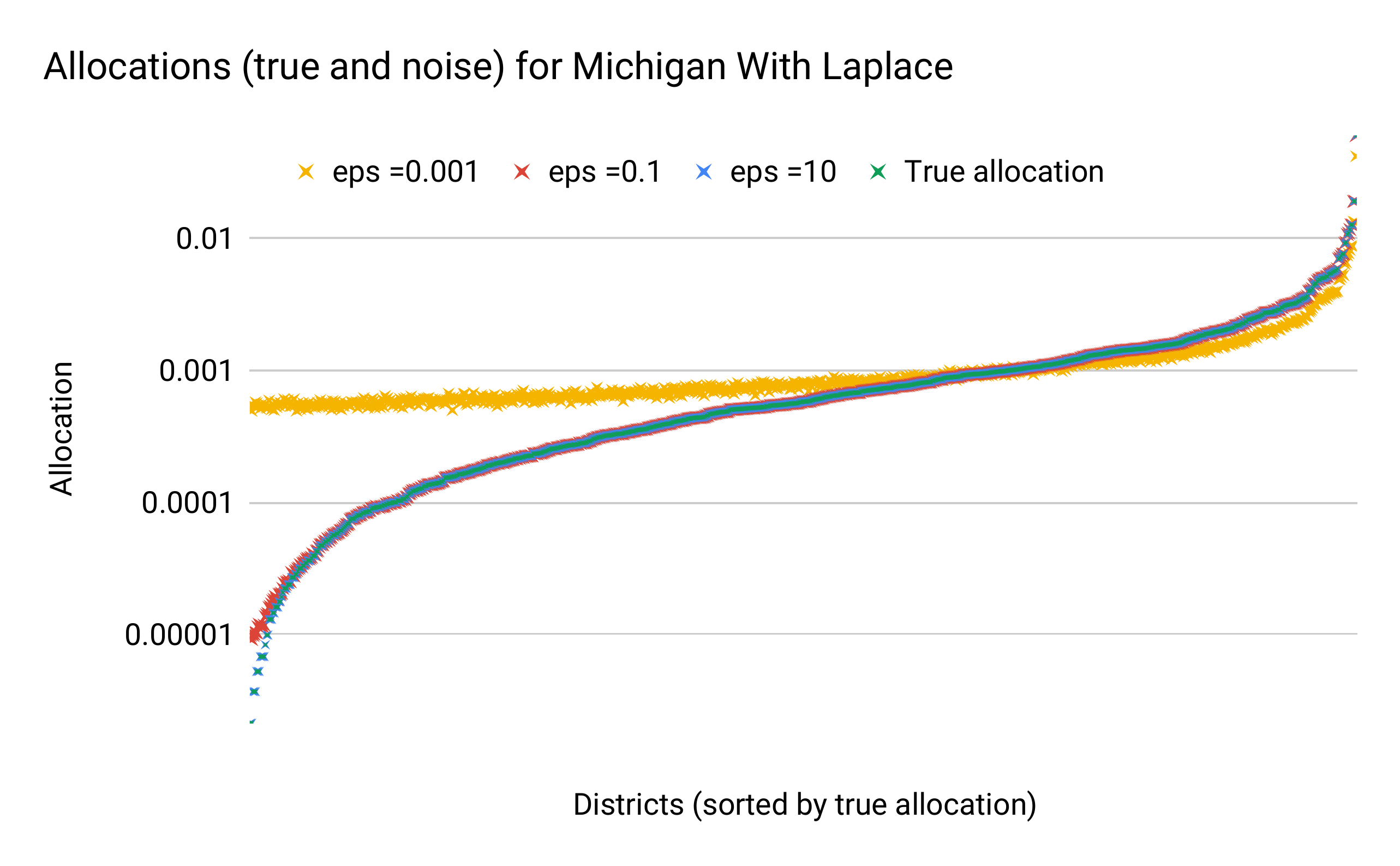}\label{fig:michigan-allocation-lap}}
	\subfigure[Multiplicative Allocation Error]{
	\includegraphics[width=0.3\textwidth]{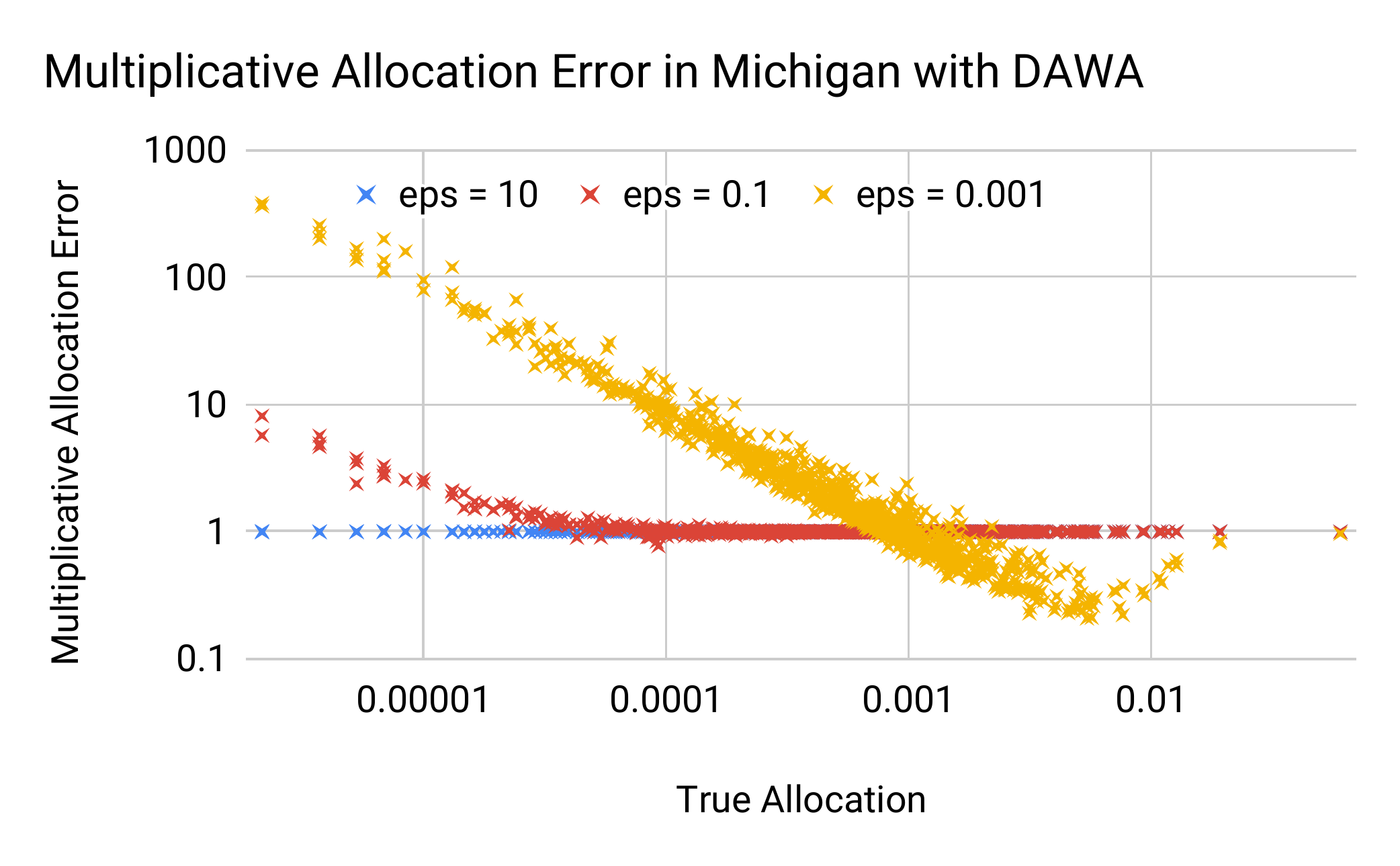}
	\label{fig:michigan-mult-dawa}}
	\subfigure[Misallocation]{
	\includegraphics[width=0.3\textwidth]{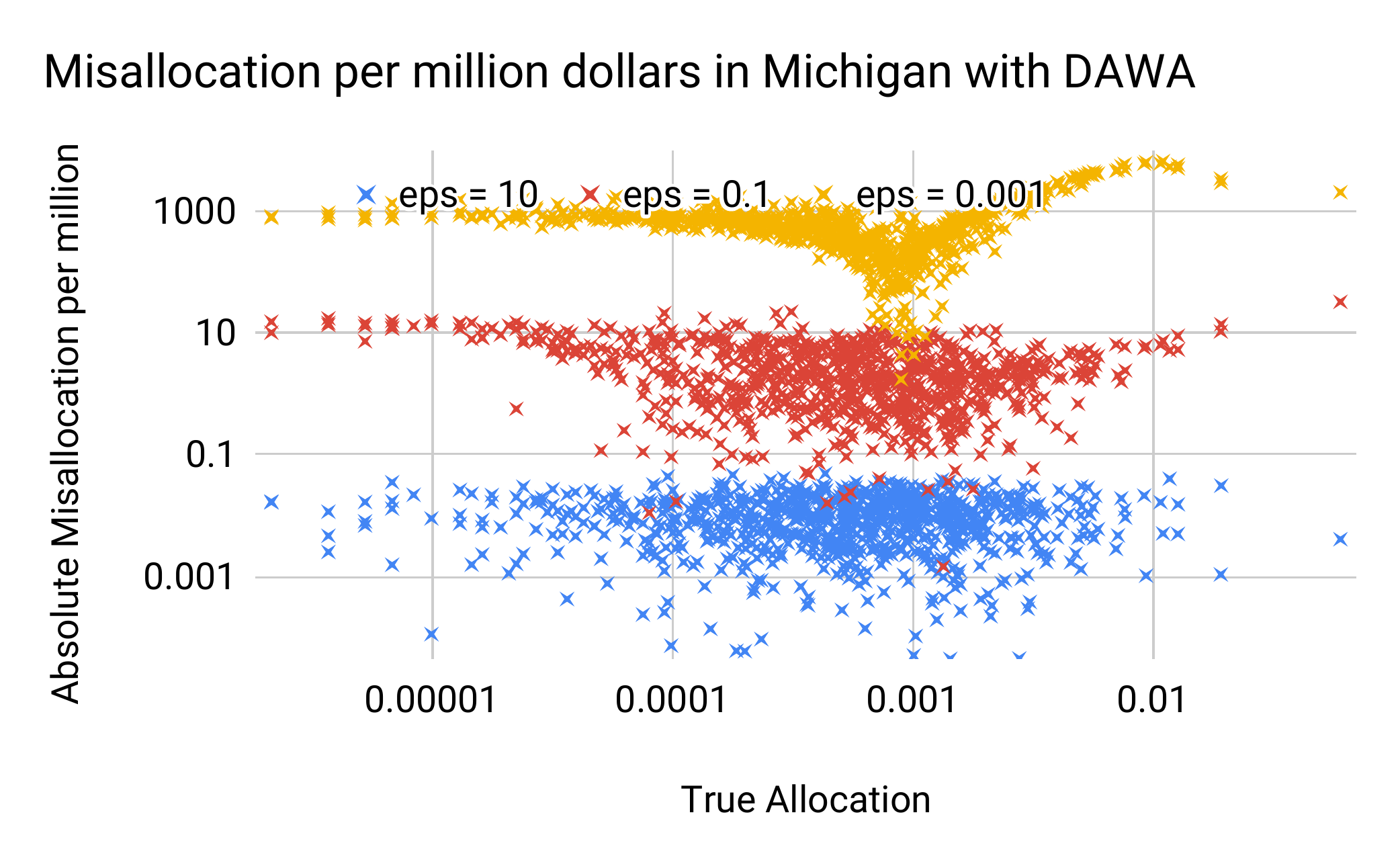}\label{fig:michigan-misallocation-dawa}}
	\subfigure[Allocations]{
	\includegraphics[width=0.3\textwidth]{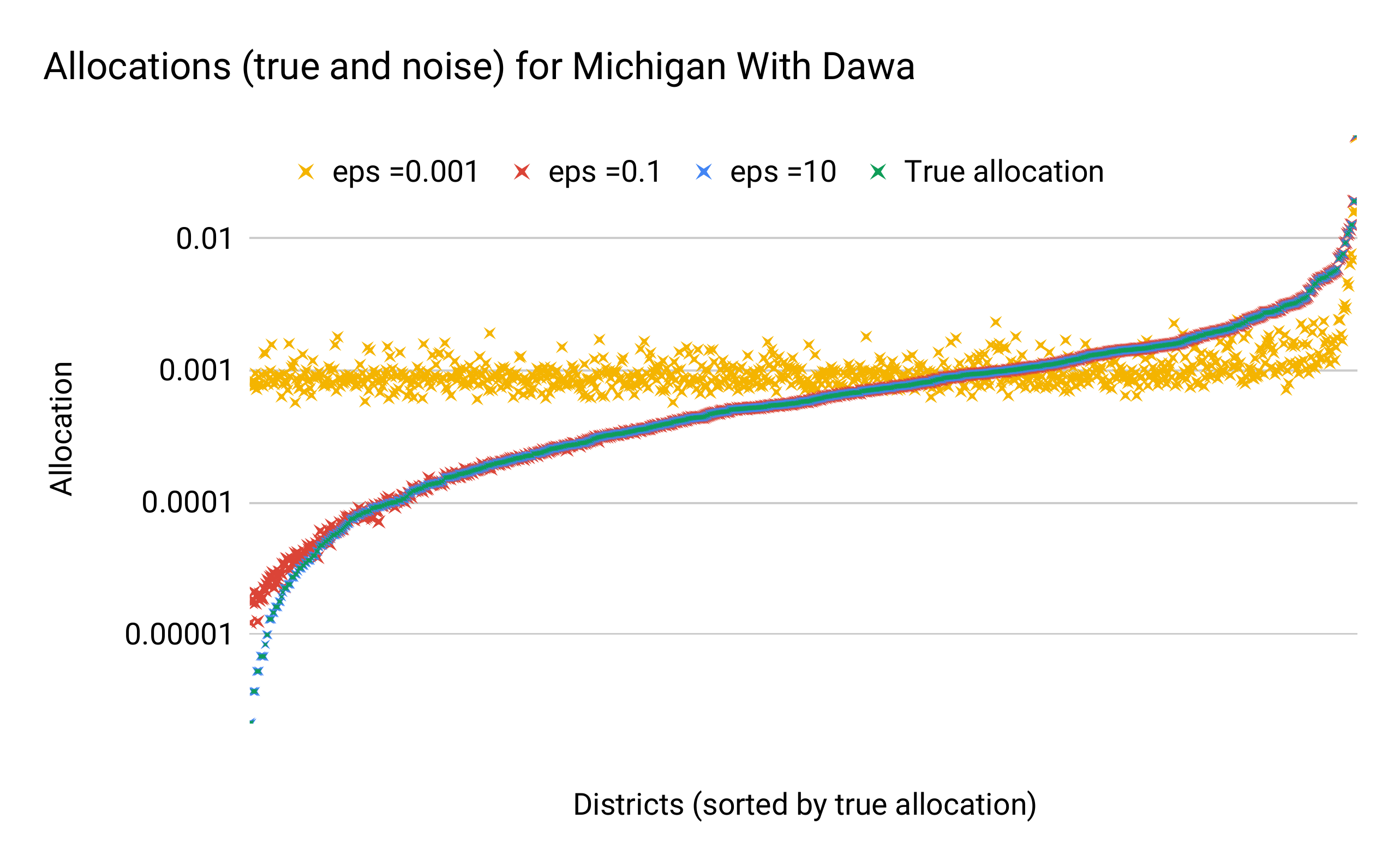}\label{fig:michigan-allocation-dawa}}
\vspace{-0.5em}
\caption{\label{fig:michigan} Fairness in allocation for Michigan using the Laplace Mechanism (top) and DAWA (bottom).}
\vspace{-0.5em}
\end{figure*}

We now turn our attention to the important class of funds allocation problems.  A recent study estimated that the annual distribution of at least \$675 billion dollars relies on data released by the Census Bureau \cite{Hotchkiss17Uses}.  This includes funding for educational grants, school lunch programs, highway construction, wildlife restoration, among many others. As an example of federal funds allocation, we consider Title I of the Elementary and Secondary Education Act of 1965 \cite{Sonnenberg16Allocating}. This is one of the largest U.S. programs offering educational assistance to disadvantaged children.  In fiscal year 2015, Title I funding amounted to a total of \$14.4 billion, of which roughly \$6.5 billion was given out through ``basic grants''  which are our focus.

\subsection{Problem Definition}
The federal allocation is divided among qualifying school districts in proportion to a count of children in the district aged 5 to 17 who live in families who fall below the poverty level or receive a form of federal financial aid \cite{Sonnenberg16Allocating}. This proportion is then weighted by a factor that reflects the average per student educational expenditures in the district's state.  The allocation formula is described formally in \Cref{tbl:titleI}, where the outcome represents the fraction of the total allocation (which changes annually) the district will receive.
%
%
\begin{table}[h]
\caption{\label{tbl:titleI} Title I Funding Allocation}
\vspace{-2ex}
\fbox{\begin{minipage}{\columnwidth}{\small
Assignees are all U.S. school districts; outcomes are the fraction of  allocated funds for each school district.
\begin{outline} \itemsep 0.5ex
\1 	Assignees: School Districts
\1  Outcome: $[0,1]$
\1 $Q = \{ exp, eli \}$ where
	\2[] $exp(I_a)$: average per student expenditures (for state containing district $a$)
	\2[] $eli(I_a)$: number of eligible students in district $a$.
\1 $\mathcal{M}(a; \mathbf{X})=\cfrac{\mathbf{X}_a^{exp} \cdot \mathbf{X}_a^{eli}}{\sum_{b \in A}\mathbf{X}_b^{exp} \cdot \mathbf{X}_b^{eli}}$
\end{outline}}
\end{minipage} }
\end{table}

\stitle{Assessing Fairness}
To assess fairness, we consider the difference between the allocation vector based on the noisy statistics $\mathbf{\tilde o}$ and the allocation vector based on true counts $\mathbf{o}$, assessing disparities across assignees (in this case, districts).  An allocation mechanism is fair if the distance measures do not vary much across districts.  We can measure fairness \textit{ex ante}, i.e., before running the (randomized) allocation mechanism, as well as, \textit{ex post}, i.e., on the outcome of the allocation. We focus on \emph{ex ante} measures as they capture disparities due to the randomized allocation mechanism.

\textbf{Multiplicative Allocation Error:} For each district $a$, we compute $\mathbb{E}[\mathbf{\tilde o}_a]/\mathbf{o}_a$. Differences in this measure across districts can be interpreted as a measure of envy or unequal treatment. For instance, an example of an unfair allocation would be one where some districts have a ratio much larger than 1, while others have a ratio smaller than 1.  In plots we show the entire distribution of the multiplicative allocation error across districts.

\textbf{Misallocation per million dollars:} For each district $a$, we also measure the dollar amount that is under or over-allocated to each district, per million dollars allocated in total: $\gamma(a) = (\mathbb{E}[\mathbf{\tilde o}_a] - \mathbf{o}_a) \cdot 10^6$. A significant difference in this measure between two districts ($\gamma(a) - \gamma(a')$) would suggest that districts are not treated equally and could be interpreted as a measure of envy. Again, in plots we show the distribution of $\gamma(\cdot)$ across all districts.

\begin{figure*}[t]
	\centering
	\subfigure[Multiplicative Allocation Error]{
		\includegraphics[width=0.3\textwidth]{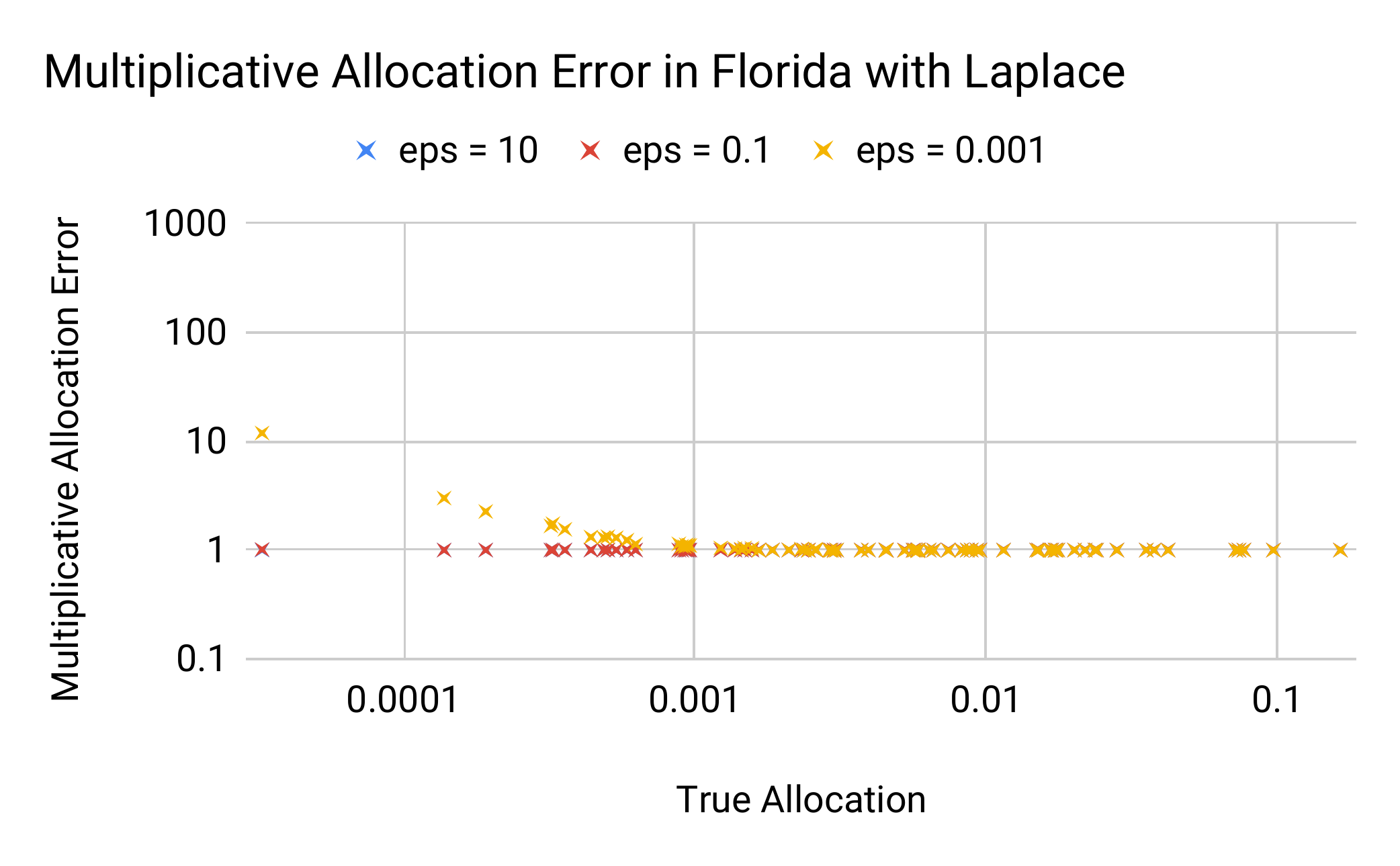}
		\label{fig:florida-mult-lap}}%
	\subfigure[Misallocation]{
		\includegraphics[width=0.3\textwidth]{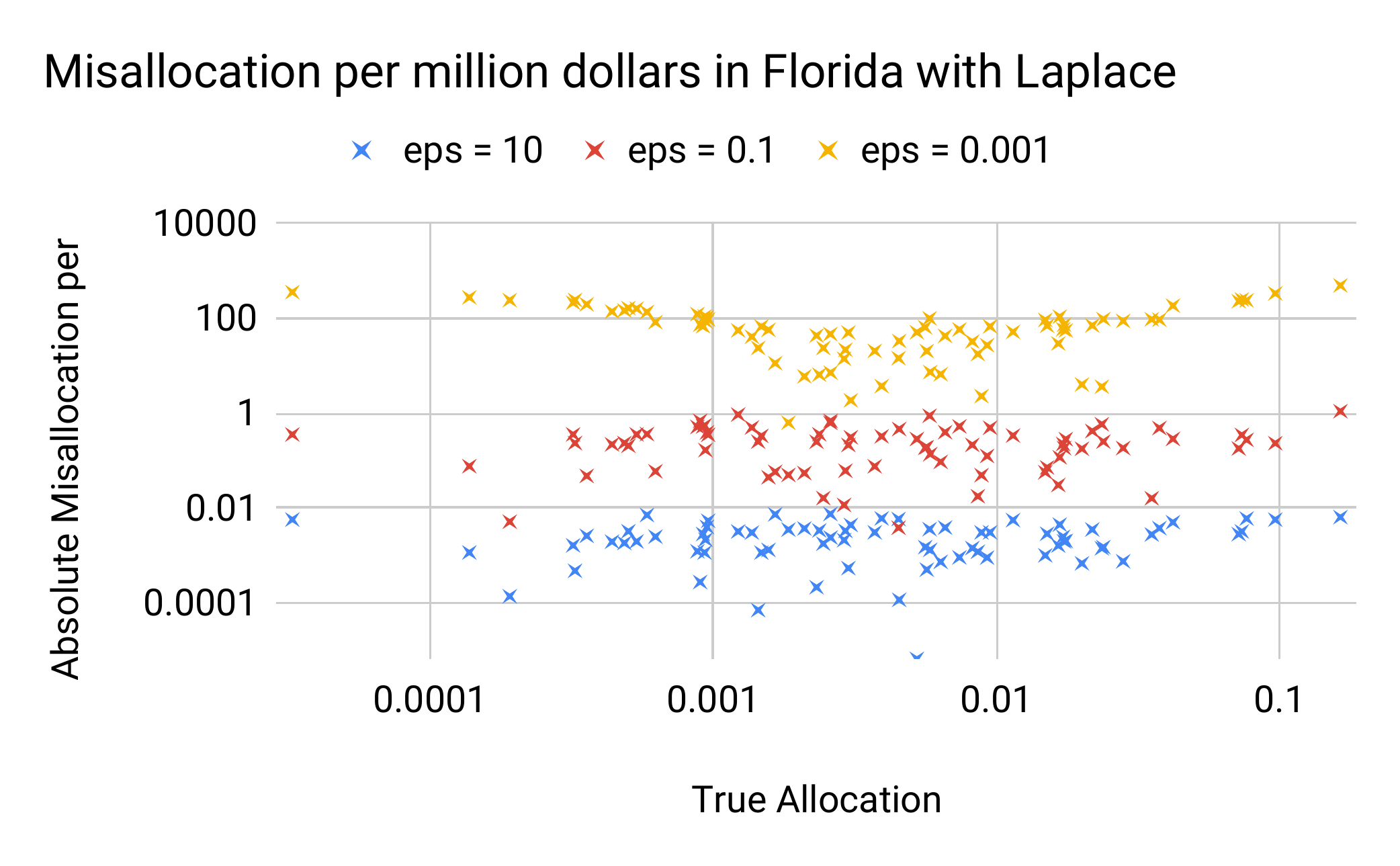}\label{fig:florida-misallocation-lap}}%
	\subfigure[Misallocation per million dollars for Michigan]{
		{\small
		\begin{tabular}[b]{| l | l | r | c |c|}
			\hline
			$\epsilon$ & Algorithm & Total  & Min & Max \\
			\hline
			\multirow{3}{2em}{$0.1$} & Laplace & 1,606 & -32& 16\\
			  & DAWA & 3,299 & -32 &22\\
			  &Inflationary & 81,887 & 84 & 99 \\
			\hline
			\multirow{3}{3em}{$0.001$} & Laplace & 563,960 & -31,137& 837  \\
			 & DAWA & 741,085 & -9,051& 1,673\\
			 &Inflationary & 5,715,476 & 526 & 6959 \\
			\hline
		\end{tabular}
		\label{tab:stats}}}
\vspace{-0.5em}
	\caption{\label{fig:florida} Fairness in allocation for Florida with the Laplace mechanism, and misallocation statistics for Michigan.}
\vspace{-0.5em}
\end{figure*}

\subsection{Empirical Findings}
\stitle{Experimental Setup}
The exact counts of Title I eligible students per district are unavailable so,  as a proxy, we used per-district counts of free-lunch eligible students as reported by
The National Center for Education Statistics for years 2013-2014.
For simplicity, we treat the average per student expenditures $exp(I_a)$ as public, following \cite{Abowd18Economic}.  We obtained data for 15650 of 18609 school districts.
%
%
We use two differentially private algorithms to estimate $eli(I_a)$ for each $a$: the Laplace mechanism and DAWA. The former adds independent 0-mean noise to the count in each district. The latter adds noise to the total count of groups of districts rather than individual districts. The total noisy count of a group of districts is then evenly divided among districts in the group. 
In both algorithms, negative counts are rounded to zero. The resulting vector of student counts may be fractional, but it is non-negative. 

\eat{the Laplace mechanism to estimate $eli(I_a)$ for each $a$, rounding negative counts to zero.
The resulting vector of student counts may be fractional, but it is non-negative.  We also use the \textit{DAWA} algorithm \cite{Li14Data-}, as described in \Cref{sec:background}. \textit{DAWA} has lower variance (noise is added to the total count of district groups rather than individual groups) but may incur some bias (as districts in a group are assumed to have equal counts).  We run 1,000 trials for each algorithm, varying $\epsilon$ in a range of $10^{-6}$ to $10$.}

For clarity of presentation, we show results on two states: Michigan and Florida (see Figures~\ref{fig:michigan} and \ref{fig:florida}). We chose these states because the histograms of the number of eligible students per district show contrasting properties. We obtained data for 888 districts in Michigan, which included a number of small districts with the smallest containing just 8 eligible students. On the other hand, Florida has a smaller number of comparatively larger districts (we obtained data for 74, the smallest
having 49 eligible students).


\stitle{Finding T1:} \emph{In cases of low $\epsilon$ there are significant disparities in  outcomes (over- and under-allocation) using private statistics.}  Using the Laplace mechanism, the mean allocation for small districts is typically much higher than the true allocation while the mean allocation of larger districts is typically lower. This is shown in \Cref{fig:michigan-mult-lap}, which plots the multiplicative allocation error of a district versus its true allocation. The districts are shown sorted by true allocation. The smallest districts see a $1.01\times$ increase for $\epsilon = 10$, a 10$\times$ increase for $\epsilon = 0.1$ and a 500$\times$ increase for $\epsilon = 0.001$. The largest districts see their allocations decrease by 0.001\% for $\epsilon = 10$,  0.05\% for $\epsilon = 0.1$ and 50\% for $\epsilon = 0.001$.

The Laplace mechanism adds 0-mean noise to the data, and, in expectation, the noisy counts should be the same as the true counts. However, these counts could be negative and since negative counts are rounded to 0, this adds an upward bias to the noisy counts. Moreover, this bias increases the total number of students, thus bringing down the weight of larger districts.

 \Cref{fig:michigan-misallocation-lap} shows the absolute dollars misallocated per million dollars allocated.  In terms of raw dollar amounts, the largest districts see the greatest misallocation and see a drop in funding of about 31,000 (see \Cref{tab:stats}). On interpretation of this behavior is that larger districts are being \textit{taxed} to ensure that students in all districts enjoy the same level of privacy protection.

The results for DAWA (\Cref{fig:michigan-mult-dawa,fig:michigan-allocation-dawa}) have more disparity than those of the Laplace mechanism. At $\epsilon = 0.001$ some districts get about 555$\times$ their true allocation, while others get only a tenth of their true allocation, in expectation whereas, under the Laplace mechanism, every district gets at least 0.48x of their
true allocation.

For districts in Florida (see \Cref{fig:florida-mult-lap} and \Cref{fig:florida-misallocation-lap}), we see almost no difference at $\epsilon = 10$. At $\epsilon = 0.1$ there is very little difference between the true and noisy allocations between districts both additively and multiplicatively. At $\epsilon = 0.001$, we see the same effect of larger districts being taxed. However, the effects are less prominent than in Michigan. This is because there are fewer small counts in Florida as well as fewer districts  overall, resulting in a lower variance estimate of the total count used in the denominator of the allocation formula.

\stitle{Finding T2:} \emph{Populations with small entitlements, relative to the privacy parameter $\epsilon$, will experience significant misallocation.}  Detecting small counts or the presence of small effects in data is incompatible with differential privacy.  This is a fundamental property of any differentially private mechanism and the meaning of ``small'' depends on $\epsilon$: \emph{Any $\epsilon$-differentially private algorithm can not distinguish between counts that differ in $\tau(\epsilon, \delta) = \frac{1}{\epsilon} \log \left(\frac{1}{\delta}\right)$, with probability $1-\delta$.}  Thus, no matter what differentially private algorithm one uses, districts with sufficiently small counts will undergo mis-allocation.  Due to rounding, they tend to get higher allocations than they deserve, in expectation, at the cost of larger districts.
\abowd{Sampling error has the same problem. You might want to quantify it, too, using a parameter for the sampling rate the lies in (0,1].}

 This phenomenon is evident in \Cref{fig:michigan-allocation-lap} and \Cref{fig:michigan-allocation-dawa} which show the true and noisy allocations for all districts, when Laplace and DAWA are used respectively.  At $\epsilon =0.001$, for both mechanisms, all districts with a true allocation less than 0.001 end up with an allocation of roughly $0.001$ in expectation. This is because, in these cases, these mechanism can not distinguish between the number of students in those districts and 0, and rounding induces a positive bias.  On the other hand, the noisy allocations at $\epsilon = 0.1$ track the truth more closely (although even at $\epsilon = 0.1$, there is a threshold under which noisy counts cannot reflect their true magnitude), and at $\epsilon = 10$ the true and noisy allocations barely differ.


\stitle{Finding T3:} \emph{Under some privacy mechanisms, districts with a greater entitlement can receive a smaller expected allocation.} Consider two districts, $a$ and $b$ where $a$ has a smaller number of eligible students than $b$.  Naturally, the true allocation of $a$ will be smaller than the true allocation of $b$, and the inversion of this relationship would violate a commonly held notion of fairness.

Under the Laplace mechanism, in expectation, we can show that the allocation for $a$ will be no larger than the allocation for $b$.  However, this is not true for the DAWA algorithm, because of bias in estimated counts.  In particular, for DAWA, a smaller district may be grouped with other larger districts, while a larger district may be grouped with other smaller districts. This results in the smaller district getting a larger expected allocation than the larger district. Empirically we find that, using DAWA with $\epsilon = 0.001$, $381$ out of the $888$ districts exhibit at least one inversion, where a larger district gets a smaller allocation.

\subsection{Mitigating unfairness} \label{sec:sub:title1repair}

Here we introduce a post-processing step designed to mitigate the inequities present due to the noise introduced for privacy.  We design the approach with the Laplace mechanism in mind, but leave extensions to other mechanisms as future work.  


The goal of this method is to ensure that, with high probability, each district receives an allocation at least as large as its true allocation.  More specifically we aim to satisfy the following condition:
\begin{definition}[No-penalty allocation]\label{def:repair}
	  Given  $\mathcal{M}(B, \mathbf{x})$,
		a no-penalty allocation is any (randomized) allocation $\mathcal{M}'$ allocating a new budget $B'$ such that for all $a$; $\mathcal{M}'(a; \tilde{\mathbf{X}}) \geq \mathcal{M}(a; \mathbf{X})$  with failure probability no greater than $\delta$.
\end{definition}
We propose a repair mechanism that achieves the above definition, but requires inflating the overall allocation in order to guarantee (with high probability) that no district is disadvantaged. In particular, our inflationary repair algorithm inflates the counts of each district by a slack variable $\Delta =  \ln(2k/\delta)/\epsilon$ while deflating the total count of all districts by another slack variable $\Delta'  = k\ln(2k^2 / \delta)/\epsilon$ (where $k$ is the total number of districts).  The final allocation is:

$$ \mathcal{M}'(a; \tilde{\mathbf{X}})=\cfrac{\mathbf{X}_a^{exp} \cdot \tilde{\mathbf{X}}_a^{eli} + \Delta}{\sum_{b \in A}\mathbf{X}_b^{exp} \cdot \tilde{\mathbf{X}}_b^{eli} - \Delta ' } $$
Note that both $\Delta$ and $\Delta'$ depend on $\epsilon$ as they are calibrated to the added noise.
We prove in \cref{suppl:title1} that, for any given $\delta$, this algorithm provides a no-penalty allocation.

\begin{theorem} \label{theorem:repair}
	The repair algorithm satisfies \cref{def:repair}.
\end{theorem}

We used this inflationary allocation method in the Title I funds allocation experiment described above, setting the acceptable failure probability, $\delta=.05$.  Compared with the standard allocation applied to the private counts resulting from the Laplace mechanism, it removes penalties experienced by some districts, at the cost of inflating the overall allocation.
For $\epsilon = 0.1$, in expectation, the repair mechanism requires increasing the budget by a relatively modest factor of $1.082 \times$ the original budget (i.e. $\$82,000$ per million).  At lower levels of epsilon, achieving a no-penalty allocation has a significant cost.  At $\epsilon = 0.001$, in expectation,  the mechanism allocates $5.715 \times$ the original budget.  These results are included in \cref{tab:stats} and can be compared with the expected misallocation under the Laplace or DAWA privacy mechanisms when combined with the standard allocation rule.

This repair algorithm mitigates one aspect of unfairness, since districts cannot complain that they were disadvantaged as a result of the noise added to population counts to preserve privacy.  Of course, there may still be inequity in the allocation, since some districts' allocations can be inflated more than others.

This approach to mitigation does not alter the underlying privacy mechanism, which may be seen as an advantage to a statistical agency publishing privatized counts to be used for many purposes beyond funds allocation.  However, the proposed algorithm relies on an analysis of the noise distribution.  While feasible for the noise introduced by the Laplace mechanism, it would need to be adapted to a mechanism like DAWA whose error distribution is data-dependent, possibly requiring additional consumption of the privacy budget to calibrate the slack variables.

\section{Problem 3: Apportionment of Legislative Representatives} \label{sec:apportionment}

Apportionment is the allocation of representatives to a state or other geographic entity.
We use parliamentary apportionment as our example domain, and consider the particular case of allocating representatives to the Indian Parliament's Lower House, in which a \emph{fixed} number of representatives ($543$) are apportioned among $35$ states,\footnote{We use
'state' to refer to both states and union territories} with each state having at least one representative.

Parliamentary apportionment is carried out using population counts obtained from a census.
While state population counts are aggregates over large groups, they nevertheless cannot be released in unmodified form in the standard model of differential privacy (i.e., without an infinite privacy loss parameter), as they could reveal presence of individuals in combination with other statistics.

%

In experiments, we consider $\epsilon$ values and requisite noise sufficient to impact apportionment methods.  Whether or not this degree of noise would be used in practice for congressional apportionment, the findings
apply
to apportionment problems over smaller geographies (e.g., allocating seats on a school board to school districts).  In particular, Laplace noise required to provide privacy at $\epsilon=10^{-4}$ on a population of $10^6$ (a small state) has equivalent effects as using $\epsilon=.1$ on a smaller population of $10^3$ (a small school district).

\subsection{Problem Definition}

The principle underlying fair apportionment is equal representation. Therefore the ideal allocation of seats to a state is given by the state's \emph{quota}, which is its fraction of the population multiplied by the total number of representatives.  A state's quota is typically non-integral, but an integral number of seats must be apportioned.  Thus any selected apportionment outcome will deviate from the quota values, leading to some degree of disparity in representation.
There are various apportionment methods studied in literature~\cite{Peyton-Young04Fairness,Balinski01Fair}.
In this paper we do not make a comparison between these algorithms, rather we are interested in how adding Laplace noise affects representation of states with different population counts.  Thus, we apply the following simple algorithm (\Cref{tbl:apportionment}):
We compute the quotas for all states and round them to the nearest integer, with the constraint that every state receives at least one seat.  This algorithm is not guaranteed to allocate exactly 543 seats.

\begin{table}[b]
\vspace{-2em}
\caption{\label{tbl:apportionment} Apportionment of seats in parliament seats}
\vspace{-1ex}
\fbox{\begin{minipage}{\columnwidth} {\small
Assignees are all Indian states; outcomes are seats in the Lower House of Parliament.
\begin{outline} \itemsep .5ex
\1 Assignees: States 
\1 Outcomes: $\{1, 2, 3, \dots\}$
\1 $Q = \{ tot \}$
	\2[] $tot(I_a)$: total population in state $a$.
\1 $\mathcal{M}(\mathbf{X}_\mathbf{a})=$
    \2 Calculate quota: $q_a = \frac{\mathbf{X}^{tot}_a}{\sum_{b \in A} \mathbf{X}^{tot}_b} \cdot 543$ 
    \2 Round to nearest positive integer: $\max \set{ Round({q_a}), 1}$   
\end{outline}}
\end{minipage}}
\end{table}


\begin{figure*}[t]
	\centering
	\subfigure[Average-expected-deviation]{
	\includegraphics[width=0.32\textwidth]{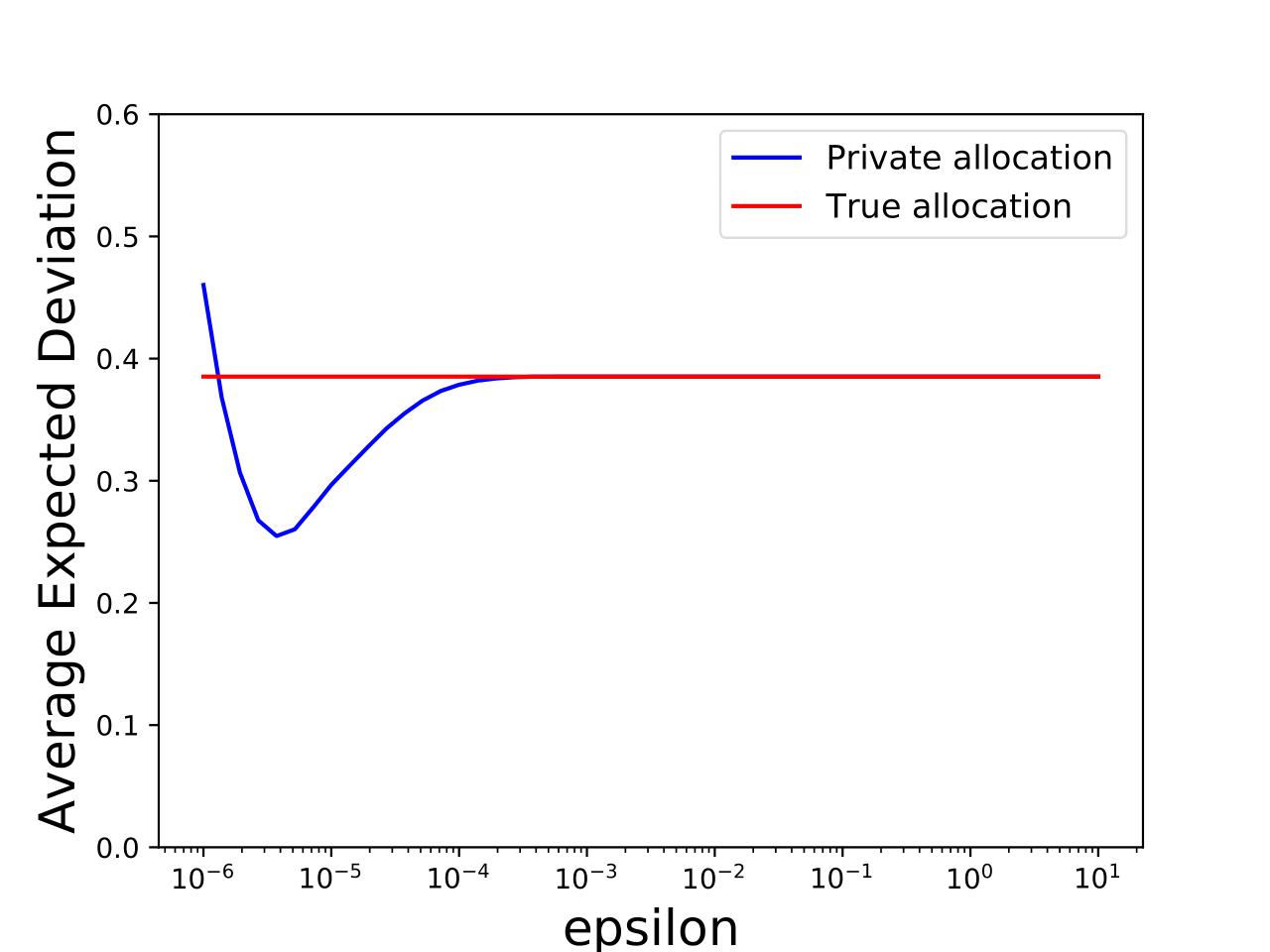}
	\label{fig:app-exante}}
	\subfigure[Per-state expected deviation ($\epsilon=1.4 \times 10^{-5}$)]{
	\includegraphics[width=0.32\textwidth]{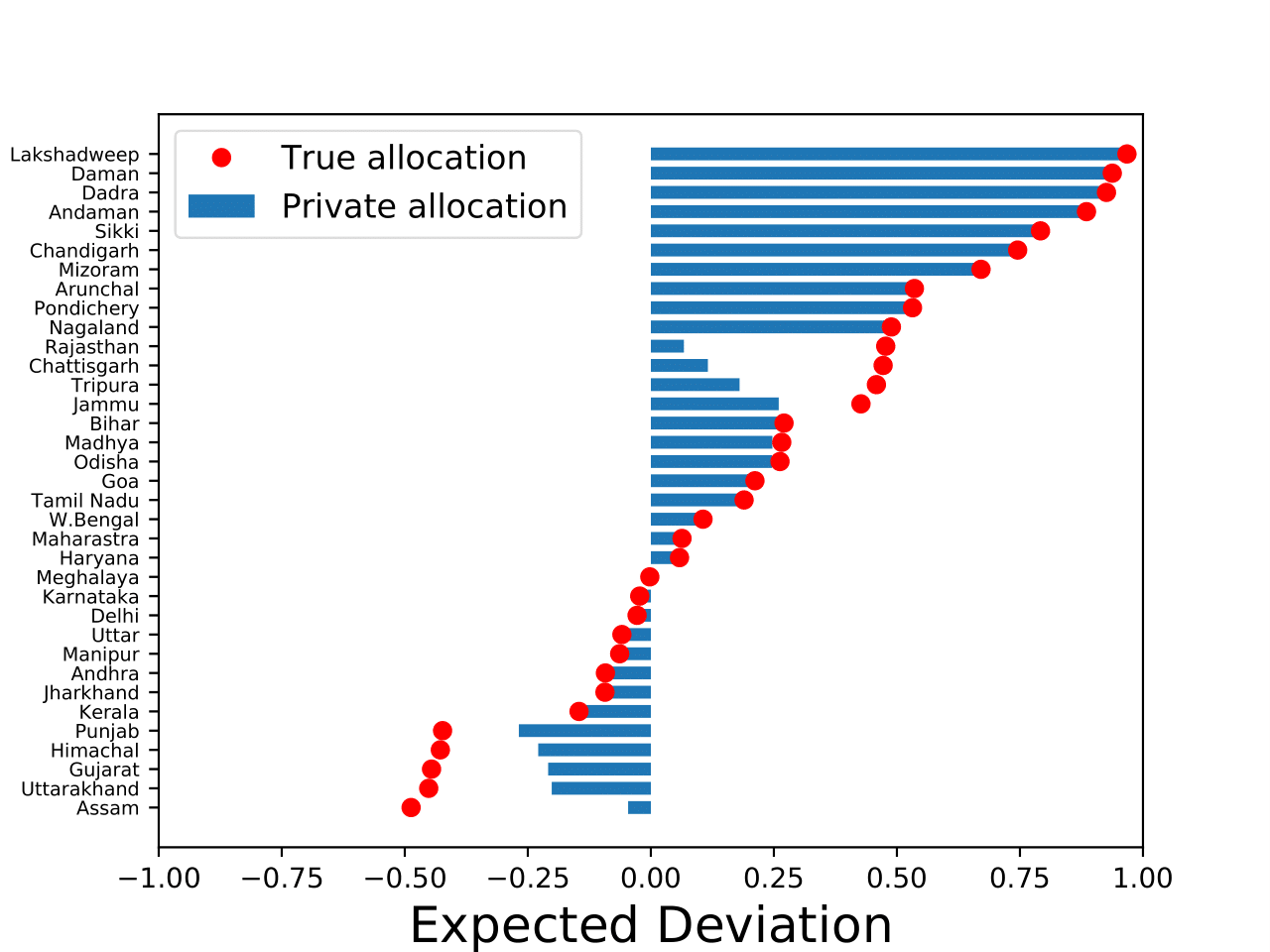}\label{fig:app-expdev}}
	\subfigure[Max-multiplicative fairness]{
	\includegraphics[width=0.32\textwidth]{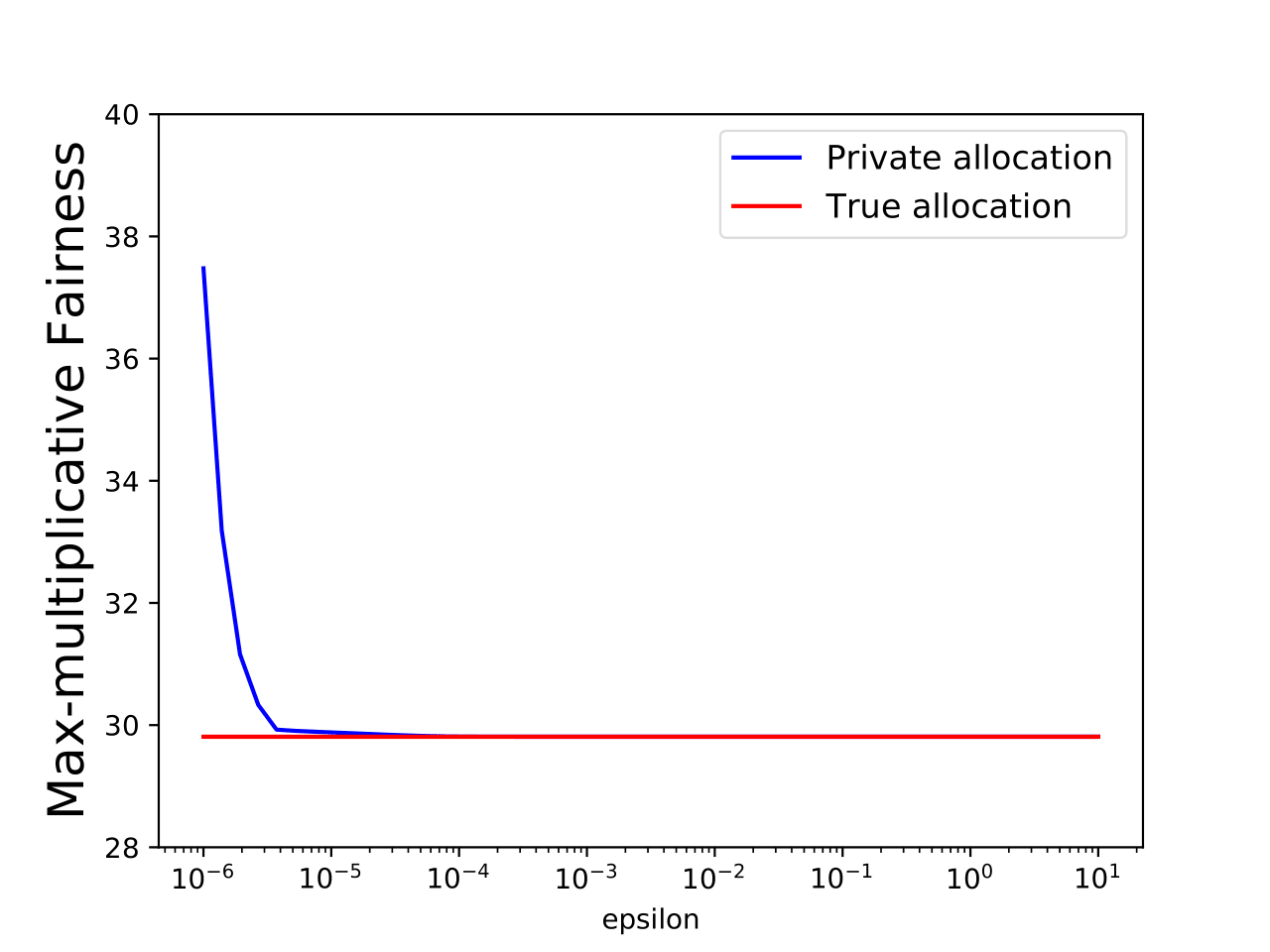}\label{fig:app-expost}}
\vspace{-0.5em}
\caption{\label{fig:apportionment} Allocation of seats to the Lower House of the Indian Parliament using population counts with Laplace noise.}
\vspace{-0.5em}
\end{figure*}

\stitle{Assessing fairness}
A desirable fairness property is \emph{quota satisfaction}; i.e., the number of seats apportioned to a state should be (roughly) proportional to the population of the state.
%
When we add Laplace noise, this property may not hold when considering specific random outcomes (i.e. ex-post), but could hold in expectation, hence we focus on the deviation from the ideal standard of equal
representation---i.e., the quota values.  We consider the following two measures, where $\mathbf{q}_a$ denotes the quota for state $a$ (computed on the true population counts):

{\bf Max-multiplicative}: this measure considers pairs of states and quantifies the disparity between the ratio of their allocation and their quota:
$\mathbb{E}\Big[\max_{a,b \in A} \Big\{\frac{\mathbf{\tilde o}_a}{\mathbf{q}_a} - \frac{\mathbf{\tilde o}_b}{\mathbf{q}_b}\Big\}\Big]$.
%
Given a particular outcome $\mathbf{o}$, this measure can be interpreted as capturing the maximum incentive for an individual to move from one state to another state in order to increase their representation.  We consider the expectation of this measure over the randomness in the privacy mechanism.

{\bf Average-Expected-Deviation}:  We also consider the expected absolute deviation from quota on a per-state basis, which we then average over the states:
$\frac{1}{|A|}\sum\limits_{a \in A} \lvert \mathbb{E}[\mathbf{\tilde o}_a]-\mathbf{q}_a \rvert$.
When this measure is small, it means that most states will receive, on average over the long run, an apportionment close to their quota.

These measures are quite different, as our empirical results will show.  The first is based on an \emph{ex-post} measure of fairness, which can be evaluated on a single apportionment outcome; we consider the expected value of this measure.  The second isolates a particular state, evaluating the difference from quota of the expected apportionment for that state, and then aggregates over the states.  It can
be seen as an \emph{ex ante} measure of fairness: if, for example, two states had equal expected deviation from quota, then, prior to any execution of the randomized algorithm, they may not prefer the other state's future outcome.  We note that an expected deviation from quota of zero was used as a fairness criterion by Grimmett in the context of a randomized (but non-private) method for apportionment~\cite{Grimmett04Stochastic}.


\vspace{-0.5em}
\subsection{Empirical Findings}

\vspace{-0.5em}
\stitle{Experimental Setup}  We used the 1971 state population totals published by the Indian Parliament in the budget of 2006/07
, which provides data for 35 states and union territories. 
\cite{indiacensus}
We evaluate the impact on apportionment outcomes when state population totals are perturbed by the Laplace Mechanism, for varying $\epsilon$.  We do not consider more sophisticated privacy mechanisms (as we did earlier) because, for this small set of statistics, they do not improve upon the Laplace mechanism.

\stitle{Finding A1.} {\em For some $\epsilon$, noise introduced into population totals can lead to more fair apportionment, in expectation.}
\Cref{fig:app-exante} shows the average-expected-deviation measure as it varies with $\epsilon$.  We see that the introduction of noise actually improves over the baseline deviation from quota, between approximately $\epsilon=1.4 \times 10^{-6}$ and $\epsilon=3.8 \times 10^{-4}$.
This is because randomization can reduce, on average, the deviation from quota caused by the integrality constraint.

A more detailed look
is provided by \Cref{fig:app-expdev}, which shows per state results for a single privacy level,
$\epsilon=1.4 \times 10^{-5}$.
For each state, the red dot shows the deviation from quota on the true population totals (which may be positive or negative).  The blue bars show the expected deviation from quota for the respective state, often
substantially lower.
While this decreased deviation is interesting, the expected apportionment is an unattainable outcome in any possible trial, so this may be an unsatisfying property in practice.

\stitle{Finding A2.} {\em As $\epsilon$ decreases, apportionment outcomes display a greater multiplicative disparity between most favored and least favored state.}
\Cref{fig:app-expost} shows the max-multiplicative measure as it varies with $\epsilon$ and here we see the fairness measure worsen as noise increases.  When considering this ex-post measure, noise does not help: apportionment outcomes tend to include states receiving substantially more than their quota while others receive substantially less, and the disparity increases with the magnitude of the noise.

\section{Conclusion} \label{sec:conclusion}
We empirically measure the impact of differentially private algorithms on allocation processes, demonstrating with important practical examples that  disparities can arise, particularly for smaller, more protective values of the privacy-loss budget.  Some practical deployments of differential privacy have been revealed to use high privacy-loss budgets \cite{Greenberg16Apples}, which would diminish impacts, however, we emphasize that the privacy loss budget must cover {\em all} public releases, including the supporting statistics of any required allocation problems.  Thus, in practice, the privacy loss budget devoted to the statistics for any single allocation problem may be small.

The disparities in outcomes have multiple causes, including bias added by some privacy algorithms, threshold conditions inherent to some decisions, and divergent treatment of small and large populations.  Our results show that designers of privacy algorithms must evaluate the fairness of outcomes, in addition to conventional aggregate error metrics that have historically been their focus.

We proposed remedies to the disparities demonstrated for funds allocation and voting benefits, but further investigation of mitigating technology is needed.  One potential approach is to customize privacy mechanisms, targeting performance on specific assignment problems.  While this approach should be pursued, it presents  agencies like the Census with the difficult prospect of designing an algorithm for each of the thousands of assignment problems that rely on the public data they release.  Our remedies adapted allocation methods to account for the noise added by a version of the Laplace mechanism.  But some algorithms (including DAWA) do not directly support the release of error bounds, confounding this approach. Furthermore, modifying allocation procedures could be inconsistent with the governing regulations.  We hope to continue to develop and evaluate these approaches in future work.

\vspace{1ex}
{
\small
\textbf{Acknowledgments}
We are grateful for helpful discussions with Census Bureau staff, including John Abowd. This work was supported by the National Science Foundation under grants 1741254, 1409125; and by DARPA and SPAWAR under contract N66001-15-C-4067. The U.S. Government is authorized to reproduce and distribute reprints for Governmental purposes notwithstanding any copyright notation thereon. The views, opinions, and/or findings expressed are those of the author(s) and should not be interpreted as representing the official views or policies of the Department of Defense or the U.S. Government.

}

\newpage
\bibliographystyle{abbrv}
\bibliography{fair}

\section{Appendix} \label{sec:supplement}

The following appendices provide additional algorithm background and a proof omitted from the body of the paper.

\subsection{Algorithm background}

The privacy algorithms used in experiments were implemented using the Ektelo framework~\cite{ektelo}, which is available open-source.\footnote{\url{https://ektelo.github.io}} The main algorithms used were described in \cref{sec:background} and in the respective sections where they were adapted to specific problems.  We include additional background and details to aid intuition about algorithm performance and to support reproducibility.



\subsubsection*{Laplace Mechanism}

Recall that the Laplace mechanism ~\cite{Dwork06Calibrating} adds noise sampled from a mean-zero Laplace distribution.  The scale of the noise is calibrated to $\epsilon$ and a property of the computed quantity called the sensitivity.  We use a variant of the Laplace mechanism in which the desired statistics are expressed in vector form and the sensitivity is calculated automatically~(cf. VectorLaplace in ~\cite{ektelo}).  We review the uses of the Laplace mechanism for each problem domain:

\begin{description}

\item[Title I] The sensitivity of the collection of statistics used in the Title I allocation is 1 and we use a direct application of the Laplace mechanism, followed by simple post-processing consisting of setting negative counts to zero. 

\item[Voting Rights] 
For the voting rights benefits, we use a slight adaptation of the Laplace Mechanism, which we call \ident. Applying the standard Laplace Mechanism to the original queries $Q = \{ vac, lep, lit \}$ would require noise scaled to a sensitivity of 3.  Instead, we used the \ident algorithm, which adds noise to decomposed queries $Q'=\{q_1, q_2, q_3\}$ where:
\begin{eqnarray*}
	q_1 & = & lit \\
	q_2 & = & lep - lit \\
	q_3 & = & vac - lep
\end{eqnarray*}
These queries together have sensitivity one (because the addition or removal of any individual can change only one query answer, by a value of one).  The \ident algorithm uses the Laplace mechanism to estimate answers to $Q'$ and then derives estimates to $Q$ from them.  In particular,
\begin{eqnarray*}
	{\tilde X}_{lit} & = & {\tilde q}_1 \\
	{\tilde X}_{lep} & = & {\tilde q}_1 + {\tilde q}_2\\
	{\tilde X}_{vac} & = & {\tilde q}_1 + {\tilde q}_2 + {\tilde q}_3
\end{eqnarray*}
In experiments, \ident performed consistently better than a standard application of the Laplace mechanism.

\item[Apportionment]
We used a standard application of the Laplace mechanism for this problem, applied to a set of population totals, which together have sensitivity 1. As in the Title I problem, negative population counts were rounded to zero.

\end{description}

\subsubsection*{DAWA}
The Data- and Workload-Aware Algorithm (DAWA)~\cite{Li14Data-} applies a differentially private pre-processing step to group together statistics and smooth their estimates.  It can be applied to an ordered sequence of statistics, such as a histogram, and it selects a partition of the statistics into contiguous intervals so that statistics with similar value are grouped together. It uses the Laplace Mechanism as a subroutine to measure each group, and derives smoothed estimates for the statistics within each group.  This grouping and smoothing can significantly reduce total error in some settings and was shown to outperform a number of competing mechanisms on real datasets~\cite{Li14Data-}.  

However, the benefits achieved in total error come with some added complexities for practice use.  First, noisy counts produced by the algorithm have lower mean-squared-error, but that error is not unbiased, as it would be if the Laplace mechanism was used.  In the problems we consider, the counts released correspond to population counts for geographic regions.  It is difficult to predict a priori where bias may occur when using the DAWA algorithm, but informally it tends to arise for regions that are outliers amongst their neighbors.  For example, when a district has a much higher count than its neighboring districts, and those neighboring districts share roughly uniform counts, the district will tend to have its count biased downward by the grouping and smoothing process.  A related consequence is that the expected error of a given count is not simply a function of the privacy parameters used in the mechanism, but depends on the data, and thus the expected error is itself a sensitive quantity that cannot be released unless the privacy cost of this release is accounted for.    

	The complexities inherent to the DAWA algorithm are shared by many state-of-the-art privacy algorithms, the result of increasingly complex algorithmic techniques invented by the research community to lower aggregate error.  For this reason, it is important to consider a representative algorithm of this kind and assess the impact on fairness and fairness mitigations. 

We review the uses of DAWA for each problem domain:
\begin{description}

\item[ Title 1]: We run DAWA on the same data used by the Laplace mechanism. Since the ordering of the cells can affect the output of the DAWA algorithm, and since there is no well-defined order on geographic regions, we simply ordered the districts alphabetically by district name. We then set all negative counts in the output to 0.

\item[Voting Rights] We run DAWA on the same underlying data vector used by \ident --- i.e., the vector produced by applying the queries $Q'$ to the data.  This results in a data vector of size $3\cdot 5180 = 15540$.  We ordered the cells by state id, then county id. 

\item[Apportionment] DAWA was not used in the apportionment experiments because, for small numbers of counts, such as the 35 Indian states, DAWA does not outperform the Laplace Mechanism.

\end{description}

\subsection{Proofs}\label{suppl:title1}

%

\stitle{Proof of \cref{theorem:repair}} 
For any $\delta>0$, the Inflation repair algorithm described in \cref{sec:sub:title1repair} is a no-penalty allocation.

\begin{proof}

	Let $x_1, ..., x_k$ denote the true counts and $\tilde{x_1}, \ldots, \tilde{x_k}$ be the noisy counts after applying the Laplace mechanism. Let $n$ be the sum of the true counts and $\tilde{n}$ be the sum of the noisy counts. 
	
	Let $\Delta = \frac{2\ln(2k/\delta)}{\epsilon}$.  We know the following:
	\[Pr[|\tilde{x_i} - x_i| > \Delta] < \delta/2k\]
	Therefore for the sum of k districts  $\Delta' = \frac{ k\ln(2k^2/\delta)}{\epsilon}$
	\[Pr[|n - \tilde{n}| > \Delta' ] < \delta/2k\]
	Therefore, from the union bound, 
	\[Pr[x_i (\tilde{n} - k\Delta_n) < (\tilde{x_i} - \Delta)n]< \delta/k\]
	Then again using the union bound 
	\[Pr\left[\forall i \frac{\tilde{x_i} + \Delta}{\tilde{n} - \Delta'} < \frac{x_i}{n}\right] > \delta \]
	Then take the negation 
	\[Pr\left[\forall i \frac{\tilde{x_i} + \Delta}{\tilde{n} - k\Delta'} \geq \frac{x_i}{n}\right] > 1- \delta \]
\end{proof}

\end{document}